\documentclass[12pt]{article}

\usepackage[T1]{fontenc}
\usepackage[utf8]{inputenc}
\usepackage{amsfonts,amsmath,amssymb,accents,mathrsfs}
\usepackage[retainorgcmds]{IEEEtrantools}
\usepackage{multirow}
\usepackage{multicol}
\usepackage{tikz}
\usepackage{graphicx,color}
\usepackage{booktabs}
\usepackage{placeins}
\usepackage{mathtools}
\usepackage{calc}
\usepackage[nosort,nomove,compress]{cite}
\usepackage[scaled=0.9]{helvet}
\usepackage[letterpaper,
			hmargin={1.9cm,1.9cm},
			vmargin={1.9cm,1.9cm},
			footskip=8mm]{geometry}
\usepackage{setspace}
\usepackage[colorlinks,
			breaklinks,
			hyperfootnotes,
			linktoc=page,
            linkcolor=red,
            citecolor=purple,
            urlcolor=blue,
            unicode,
            pdfstartview=FitH,
            bookmarks,
            bookmarksnumbered]{hyperref}

\renewcommand\a{{\alpha}}
\renewcommand\b{{\beta}}
\newcommand\g{{\gamma}}
\renewcommand\d{{\delta}}

\newcommand\ad{{\dot{\alpha}}}
\newcommand\bd{{\dot{\beta}}}
\newcommand\gd{{\dot{\gamma}}}

\newcommand\Ysf{{\textsf{Y}}}
\newcommand\N{{\mathcal{N}}}
\newcommand\E{{\mathcal{E}}}
\newcommand\I{{\mathcal{I}}}
\newcommand\K{{\mathcal{K}}}
\newcommand\J{{\mathcal{J}}}

\newcommand\D{{\rm D}}
\newcommand\Dd{{\bar{\rm D}}}
\newcommand\pa{{\partial}}

\def\bea{\begin{IEEEeqnarray*}}
\def\eea{\end{IEEEeqnarray*}}
\def\be{\begin{eqnarray}}
\def\ee{\end{eqnarray}}
\def\n{\IEEEyesnumber}
\def\sn{\IEEEyessubnumber}

\allowdisplaybreaks

\makeatletter
\renewcommand\section{\@startsection{section}{1}{\z@}
              {3ex plus-1ex minus-.2ex}{1pt plus1pt}
              {\large\sf\bfseries\boldmath}}
\renewcommand{\subsection}{\@startsection{subsection}{2}{\z@}
              {1.5ex plus-1ex minus-.2ex}{0.01pt plus1pt}{\sf\slshape}}
\renewcommand{\subsubsection}{\@startsection{subsubsection}{3}{\z@}
              {1.5ex plus-1ex minus-.2ex}{0.01pt plus0.2pt}{\sf\boldmath}}
\renewcommand{\paragraph}{\@startsection{paragraph}{4}{\z@}
              {.75ex \@plus.5ex \@minus.2ex}{-2mm}{\sf\bfseries\boldmath}}
\makeatother

\newcommand{\Title}[1]{ {\large \bf #1 \vspace{3ex}} }
\newcommand{\Author}[3]{ {\large #1\footnote{\href{mailto:#2}{#2}}$^{#3}$} }
\newcommand{\Inst}[2]{ \emph{\centering $^{#2}$#1} }
\newcommand{\Abstract}[1]{ { ABSTRACT}\\ [4mm]
  \parbox{142mm}{\parindent=2pc\indent\baselineskip=14pt plus1pt #1}}

\begin{document}
\thispagestyle{empty}
\vspace*{6mm}
\begin{center}
\Title{Superspace formulation of massive half-integer superspin} \\   [9mm]
\Author{Konstantinos Koutrolikos}{konstantinos\_koutrolikos@brown.edu}{a,b}\\ [8mm]
\Inst{Brown Theoretical Physics Center,\\[1pt]
      Box S, 340 Brook Street, Barus,
      Providence, RI 02912, USA}{a}\\[8pt]
\Inst{Department of Physics, Brown University,\\[1pt]
      Box 1843, 182 Hope Street, Barus \& Holley,
      Providence, RI 02912, USA}{b}\\ [10mm]
\Abstract{An explicit form for the Lagrangian of a massive arbitrary half-integer superspin $\Ysf=s+1/2$
    supermultiplet is obtained in $4\D,~\N=1$ superspace. This is accomplished by the introduction
    of a tower of pairs of auxiliary superfields of increasing rank which are required to vanish
    on-shell for free theories. In the massless limit almost all
    auxiliary superfields decouple except one, which plays the role of compensator
    as required by the emergent gauge redundancy of the Lagrangian description of the massless theory.
    The number of off-shell degrees of freedom carried by the theory is
$\frac{8}{3}(s+1)(4s^2+11s+3)$. For $s=1$ our results are in agreement with those obtained
in~\cite{Gates:2013tka}. }
\end{center}
\vspace*{4cm}
\hfill{\emph{Dedicated to S. James Gates Jr. on the occasion of his 70th birthday}}\\
\vfill
%
\clearpage
%
\section{Introduction}\label{sec:intro}
Since the discovery of supersymmetry in the late 1960's and early 1970's\footnote{The dramatic story of the
discovery of supersymmetry and its early years can be found in~\cite{Kane:2000ew,Shifman:2000rs}.} there has
been tremendous progress in the development of supersymmetric theories which undoubtedly influenced a big
part of theoretical physics. For $4\D$ higher spin theories, early developments led to the supersymmetric
extension of free, massless, irreducible representations with on-shell supersymmetry and their Lagrangian
description~\cite{Curtright:1979uz,Vasiliev:1980as}. The off-shell, superspace description of such
representations was found first in~\cite{Kuzenko:1993jp,Kuzenko:1993jq,Kuzenko:1994dm}
and later developments are~\cite{Gates:2013rka,Gates:2013ska,Buchbinder:2020yip}. Furthermore, various consistent interactions
among massless higher spin supermultiplets have been discovered
recently~\cite{Buchbinder:2017nuc,Hutomo:2017phh,Koutrolikos:2017qkx,Buchbinder:2018wwg,Buchbinder:2018wzq,
Buchbinder:2018nkp,Buchbinder:2018gle,Gates:2019cnl,Metsaev:2019dqt,Metsaev:2019aig,Khabarov:2020deh}.

However, after five decades of intensive investigations, the seemingly simple question of finding the
superspace action principle for free, massive, arbitrary superspin, irreducible representation of $4\D,
\mathcal{N}=1$ super-Poincar\'{e} group still remains unanswered. This is a very important and necessary first
step in order to even consider exploring manifestly supersymmetric interactions among massive arbitrary spin
supermultiplets. The non-supersymmetric Lagrangian description of massive irreducible representations of the
Poincar\'{e} group was found in~\cite{Singh:1974qz,Singh:1974rc} almost forty years after the theory of higher
spins was first undertaken by Dirac~\cite{Dirac:1936tg}. It was demonstrated that massive higher spins carry a
very interesting and rich off-shell structure, because they require a (double) tower of auxiliary bosonic
(fermionic) fields of increasing rank.

For the case of supersymmetry such a description and understanding of the off-shell structure of arbitrary
higher spin supermultiplet is lacking, nevertheless there has been some progress towards that direction.
In~\cite{Zinoviev:2007js,Buchbinder:2019dof,Khabarov:2019dvi,Zinoviev:2020gtp,Khabarov:2020glf} a Lagrangian
description of on-shell massive higher spin supermultiplets was found. This description follows the viewpoint
of perceiving the on-shell massive spin degrees of freedom as a collection of on-shell massless
helicities\footnote{The $2s+1$ on-shell degrees of freedom of a massive spin $j=s$ can be viewed as a
    collection of the $-j,+j$ on-shell helicities for massless spins $j=0,1,\dots,s$ if $s$ is integer or
$j=1/2,3/2,\dots,s$ if $s$ is half-integer. Moreover, if one trust this analogy to hold off-shell, then one
will precisely discover the (double) tower of auxiliary symmetric (spinor) tensor fields appearing in the
off-shell Lagrangian description of~\cite{Singh:1974qz,Singh:1974rc}.} and thus, by stitching together a tower
of on-shell, massless supermultiplets with increasing spin, one can find a description for the on-shell
massive higher spin supermultiplet. However, because the supermultiplets considered only have on-shell
supersymmetry, this approach provides no clues regarding the additional auxiliary structures a manifestly
supersymmetry description may require. Manifestly supersymmetric descriptions of massive irreducible
representations do
exist~\cite{Buchbinder:2002gh,Buchbinder:2002tt,Gregoire:2004ic,Buchbinder:2005je,Gates:2006cq,Gates:2013tka}
but unfortunately not for higher spin supermultiplets.

Nevertheless, the theory constructed in~\cite{Gates:2013tka} corresponds to the massive extension of one of
the formulations of linearized supergravity which can be extended to arbitrary half-integer superspin
theories. In this paper we construct the superspace Lagrangian for a $4\D,~\N=1$ super-Poincar\'{e}
irreducible representation of mass $m$ and arbitrary superspin $\Ysf=s+1/2$. We find that in addition to the
real bosonic superfield $H_{\a(s)\ad(s)}$ which carries the propagating, on-shell degrees of freedom of this
representation, the superspace Lagrangian description requires a tower of pairs of auxiliary, fermionic
superfields $\chi_{\a(q)\ad(q-1)},~u_{\a(q)\ad(q-1)}$ with $1\leq q\leq s$. This is in agreement with the
results found in~\cite{Gates:2013tka} for the special case of $s=1$ and also is consistent with the structures
required by the non-supersymmetric results~\cite{Singh:1974qz,Singh:1974rc}. The explicit form of the
Lagrangian is found by demanding that all these auxiliary superfields vanish on-shell, in the free theory
limit and the appropriate constraints on $H_{\a(s)\ad(s)}$ are generated in order to describe the
irreducible representation $\Ysf=s+1/2$.

The paper is organized as follows. In section (\ref{sect:RevIrreps}) we review basic features of irreducible
representations of the $4\D,~\N=1$ super-Poincar\'{e} group. In section (\ref{sec:Ahis}) we construct the
superspace action that provides the off-shell description of these irreducible representations for the
arbitrary half-integer superspin supermultiplet. In section (\ref{sec:dof}) we illustrate the richness of the
off-shell structure, by counting the number of off-shell degrees of freedom of this theory.
\section[Massive and Massless Irreducible Representations]
{Massive and massless irreducible representations of $4\D$ super-Poincar\'{e}}\label{sect:RevIrreps}
Often in physics we consider some symmetry (super) group and then ask about its (irreducible) representations.
For the majority of the cases, the answer about their existence and their classification almost certainly
already exists somewhere in mathematical literature. However, for many applications in physics this is not
enough, because we do not just blindly look for any representations but for locally realizable representations
that can have a (super) field theoretic description. This tension between knowing the representation and
realize it in a field theoretic framework is responsible for the richness of many theories.

In this case we consider the $4\D,\N=1$ super-Poincar\'{e} group but in order to have a finite number of
propagating degrees of freedom we focus on its stabilizer, the super-Little group\footnote{For a detailed
review
see~\cite{Bars:1982ps,Gates:1983nr,Srivastava:1986hd,Buchbinder:1995uq,Koutrolikos:2014qjz,Gates:2013rka,
Buchbinder:2019esz}.}. The diagonalization of its two Casimir operators and its Cartan subalgebra determines
(\emph{i}) the type of superfield (number of indices and their symmetries) which realizes the
representation and
(\emph{ii}) the type of constraints it must satisfy in order to be irreducible.
For a massive, half-integer superspin $(\Ysf=s+1/2)$ representation one must consider a real bosonic
superfield $H_{\a(s)\ad(s)}$, with $s$ dotted and undotted indices which are independently symmetrized and
satisfies the following conditions
\begin{equation}\label{H}
    \Box H_{\a(s)\ad(s)}=m^2 H_{\a(s)\ad(s)}~,~\D^{\a_s}H_{\a(s)\ad(s)}=0~.
\end{equation}
These constraints kill most of the components of superfield  $H_{\a(s)\ad(s)}$ but allow the propagation of
the physical degrees of freedom of four massive spins, $j=s+1, j=s+1/2, j=s+1/2, j=s$.
Describing an irreducible representation this way is an on-shell statement, since it refers to the physical
degrees of freedom. A natural question to ask is what is the off-shell description of this representation.
This can be given in terms of an action which will generate equations of motion that give raise to the
desired constraints. In this paper, we will answer this question for the case of an arbitrary half-integer
superspin supermultiplet described by (\ref{H}).
We will show that the answer includes a hierarchy of pairs of auxiliary, fermionic, superfields
$\chi_{\a(q)\ad(q-1)},~u_{\a(q)\ad(q-1)}$ for $q=1,2,\dots,s$.

For a massless, half-integer superspin
supermultiplet, the on-shell propagating degrees of freedom are described by the superfield strength
$W_{\a(2s+1)}$. This is a chiral, fermionic superfield with $2s+1$ symmetrized undotted indices which must
satisfy the constraints
\begin{equation}
    \D^{\a_{2s+1}}W_{\a(2s+1)}=0,~\Dd_{\bd}W_{\a(2s+1)}=0~.\label{massless-constraints}
\end{equation}
This is analogous to the description of the massless spin one representation by the field
strength $F_{mn}$. It is the field strength that carries the physical degrees of freedom, since if we do an
experiment we will measure the electric and magnetic fields. $F_{mn}$ must satisfy on-shell the
constraints $\pa^{m}F_{mn}=0,~\pa_{[k}F_{mn]}=0$. One of them we call the dynamical equations of motion and the
other we call the Bianchi identity. In order to construct the off-shell, Lagrangian description
of the theory we solve the Bianchi identity in terms of a gauge vector field. The fact that the off-shell
description of the theory is given in terms of the gauge vector field, allows for a smooth transition between the
Lagrangian description of a massive spin one (Proca action) and the Lagrangian description of a massless spin
one (Maxwell's theory) by taking the massless limit of the Proca action. The discontinuity on the on-shell
degrees of freedom is recovered from the emergence of the redundancy of the vector field in the massless limit.
This approach can be applied to the case of massless irreducible representations of the
super-Poincar\'{e} group (see~\cite{Gates:2013rka}).
Briefly, the second constraint in (\ref{massless-constraints}) can be solved by expressing superfield strength
$W_{\a(2s+1)}$ in terms of a real prepotential superfield which has the same
index structure as the superfield that describes the massive theory (and thus making the transition from
massive action to to massless one smooth)
\begin{equation}
W_{\a(2s+1)}\propto\Dd^2\D_{(\a_{2s+1}}\pa_{\a_{2s}}{}^{\ad_s}\dots\pa_{\a_{s+1}}{}^{\ad_1}H_{\a(s))\ad(s)}~.
\end{equation}
The prepotential $H_{\a(s)\ad(s)}$, is now a gauge superfield because it acquires a redundancy, which leaves the superfield strength
invariant
\begin{equation}
\delta H_{\a(s)\ad(s)}=\frac{1}{s!}~\D_{(\a_s}\bar{L}_{\a(s-1))\ad(s)}
-\frac{1}{s!}\Dd_{(\ad_s}L_{\a(s)\ad(s-1))}~.\n\label{dH}
\end{equation}
Using the above redundancy as a guiding principle, one can find the superspace Lagrangian description for this
supermultiplet. The result is that there are two different off-shell formulations of the same theory based on
(\ref{dH})~\cite{Kuzenko:1993jp,Gates:2013rka,Buchbinder:2020yip}. Both
of them require the presence of an additional, unconstrained, compensating superfield. The one relevant for
our discussion\footnote{We follow the conventions of \emph{Superspace}~\cite{Gates:1983nr}.} is the
non-minimal description given by the following action principle
\begin{IEEEeqnarray*}{rl}
    S_{(\Ysf=s+1/2)}=\int d^8z~~~&H^{\a(s)\ad(s)}\D^{\g}\Dd^2\D_{\g}H_{\a(s)\ad(s)}\n\label{masslessS}\\
                                 -2~&H^{\a(s)\ad(s)}\Dd_{\ad_{s}}\D^2\chi_{\a(s)\ad(s-1)}~+c.c.\\
                                 -\frac{s+1}{s}~&\chi^{\a(s)\ad(s-1)}\D^2\chi_{\a(s)\ad(s-1)}~+c.c.\\
                                 +2~&\chi^{\a(s)\ad(s-1)}\D_{\a_s}\Dd^{\ad_s}\bar{\chi}_{\a(s-1)\ad(s)}
\end{IEEEeqnarray*}
where the unconstrained compensator has the redundancy
\begin{equation}
    \delta\chi_{\a(s)\ad(s-1)}=\Dd^2L_{\a(s)\ad(s-1)}+\D^{\a_{s+1}}\Lambda_{\a(s+1)\ad(s-1)}~.\n\label{dx}
\end{equation}
The equations of motion of the above action will generate the desired condition (\ref{massless-constraints})
and thus allow only the propagation of the helicities of spins $j=s+1$ and $j=s+1/2$.
\section{Arbitrary half integer superspin supermultiplet action}\label{sec:Ahis}
The massive extension of $S_{(\Ysf=s+1/2)}$, denoted as $S_{(m,~\Ysf=s=1/2)}$, must be such that
\begin{IEEEeqnarray*}{l}
\lim_{m\to0}\Biggl[S_{(m,~\Ysf=s+1/2)}\Biggr]=~S_{(\Ysf=s+1/2)}~+~S_{(\text{decoupled sector})}~,
\n\label{m1}\\
\d S_{(m,~\Ysf=s+1/2)}=0
\Rightarrow~\Box~H_{\a(s)\ad(s)}=m^2H_{\a(s)\ad(s)}~,~\D^{\a_s}H_{\a(s)\ad(s)}=0\n\label{m2}~.
\end{IEEEeqnarray*}
For simple cases, like the vector supermultiplet $(s=0)$ the $S_{(\text{decoupled sector})}$ is
trivially zero. However, in general we can allow such term. As an example, consider the theory of
linearized massive supergravity developed in~\cite{Gates:2013tka}. It requires the presence of an
additional auxiliary superfield $u_{\a}$ which in the massless limit decoupled from the theory. Similar
behavior is demonstrated in~\cite{Singh:1974qz,Singh:1974rc} were the massless limit of the theory generates a
non trivial decoupled sector which can be ignored from the view point of the massless theory.
Condition (\ref{m1}) dictates that the most general interaction between the auxiliary superfields in
$S_{(\text{decoupled sector})}$ and the superfields of massless theory must depend on the mass parameter.
Finally, the various numerical coefficients will be fixed by demanding that all auxiliary superfields
vanish on-shell and condition (\ref{m2}) is satisfied. This will fix the on-shell spectrum of the theory
to be the correct irreducible representation.

Motivated from the results in~\cite{Singh:1974qz,Singh:1974rc} and~\cite{Gates:2013tka}, we propose the
following coupling scheme:\\[-3mm]
\begin{figure}[ht]
    \begin{center}
    \def\svgwidth{\columnwidth}
    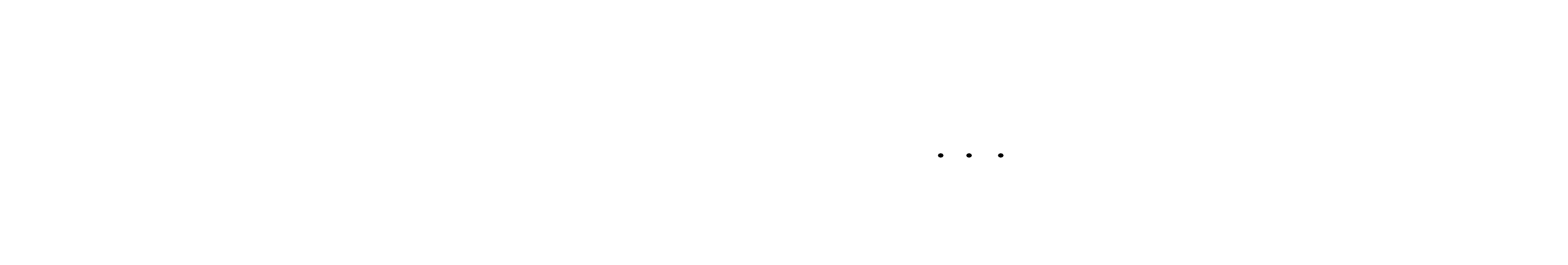
    \end{center}
\end{figure}
\\[-4mm]
Starting from the massless action $S_{(\Ysf=s+1/2)}$ we introduce a double tower of auxiliary superfields
$\chi_{\a(q)\ad(q-1)},~u_{\a(q)\ad(q-1)}$ equipped with two types of interactions. The $q$-th level superfields
$\chi^{(q)}$ and $u^{(q)}$ interact via a mass term (vertical dashed lines). This type of interaction
dissolves at the massless limit. The second type of interaction is between the $q$-th level superfield
$u^{(q)}$ and the $(q-1)$-th level $\chi^{(q-1)}$ superfield (solid diagonal lines) and is persistent, in the
sense that it survives the massless limit. Furthermore, in
compliance with the preliminary results of~\cite{Gates:2013tka}, superfield $H_{\a(s)\ad(s)}$ has its own mass
term as expected, but the auxiliary superfields $\chi^{(q)},~u^{(q)}$ do not have their own mass term.
The consistency of this scheme with the massless limit is automatic, because the first vertical dashed line
will break, decoupling the entire $\chi,~u$ sector, except $\chi^{(s)}$ and thus recovers the correct massless
action $S_{(\Ysf=s+1/2)}$.
Also it is easy to check the consistency with the $s=1$ results of~\cite{Gates:2013tka}. In that case there
is only one level $(q=1)$ and thus, only the mass interaction between $\chi_{\a}$ and $u_{\a}$ participates,
exactly as found. The most general action that reflects this coupling scheme is
\begin{IEEEeqnarray*}{rl}
    S_{(m,\Ysf=s+1/2)}=\hspace{-1mm}\int\hspace{-1mm} d^8z\Biggl\{\Biggr.
&~H^{\a(s)\ad(s)}\D^{\g}\Dd^2\D_{\g}H_{\a(s)\ad(s)}~+~m^2H^{\a(s)\ad(s)}H_{\a(s)\ad(s)}\n\label{mS}\\
&-2~H^{\a(s)\ad(s)}\Dd_{\ad_s}\D^2\chi_{\a(s)\ad(s-1)}~+c.c.\\
&-\frac{s+1}{s}~\chi^{\a(s)\ad(s-1)}\D^2\chi_{\a(s)\ad(s-1)}~+c.c.\\
&+2~\chi^{\a(s)\ad(s-1)}\D_{\a_s}\Dd^{\ad_s}\bar{\chi}_{\a(s-1)\ad(s)}\\
&\hspace{5.5cm}+m~\chi^{\a(s)\ad(s-1)}u_{\a(s)\ad(s-1)}~+c.c.\\
&\hspace{5.5cm}+c_{1}^{(s)}~u^{\a(s)\ad(s-1)}\Dd^2u_{\a(s)\ad(s-1)}~+c.c.\\
&\hspace{5.5cm}+c_{2}^{(s)}~u^{\a(s)\ad(s-1)}\D^2u_{\a(s)\ad(s-1)}~+c.c.\\
&\hspace{5.5cm}+c_{3}^{(s)}~u^{\a(s)\ad(s-1)}\Dd^{\ad_s}\D_{\a_s}\bar{u}_{\a(s-1)\ad(s)}\\
&\hspace{5.5cm}+c_{4}^{(s)}~u^{\a(s)\ad(s-1)}\D_{\a_s}\Dd^{\ad_s}\bar{u}_{\a(s-1)\ad(s)}\\
&+\sum_{q=1}^{s-1}\Bigl[\Bigr.
u^{\a(q+1)\ad(q)}\left( b_1^{(q+1)}\Dd_{\ad_q}\D_{\a_{q+1}}\chi_{\a(q)\ad(q-1)}+
b_2^{(q+1)}\D_{\a_{q+1}}\Dd_{\ad_{q}}\chi_{\a(q)\ad(q-1)}\right)\hspace{-1mm}+c.c.\\
&\hspace{9mm}+d_1^{(q)}~\chi^{\a(q)\ad(q-1)}\Dd^2\chi_{\a(q)\ad(q-1)}~+c.c.\\
&\hspace{9mm}+d_2^{(q)}~\chi^{\a(q)\ad(q-1)}\D^2\chi_{\a(q)\ad(q-1)}~+c.c.\\
&\hspace{9mm}+d_3^{(q)}~\chi^{\a(q)\ad(q-1)}\Dd^{\ad_q}\D_{\a_q}\bar{\chi}_{\a(q-1)\ad(q)}\\
&\hspace{9mm}+d_4^{(q)}~\chi^{\a(q)\ad(q-1)}\D_{\a_q}\Dd^{\ad_q}\bar{\chi}_{\a(q-1)\ad(q)}\\
&\hspace{5.5cm}+m~\chi^{\a(q)\ad(q-1)}u_{\a(q)\ad(q-1)}~+c.c.\\
&\hspace{5.5cm}+c_{1}^{(q)}~u^{\a(q)\ad(q-1)}\Dd^2u_{\a(q)\ad(q-1)}~+c.c.\\
&\hspace{5.5cm}+c_{2}^{(q)}~u^{\a(q)\ad(q-1)}\D^2u_{\a(q)\ad(q-1)}~+c.c.\\
&\hspace{5.5cm}+c_{3}^{(q)}~u^{\a(q)\ad(q-1)}\Dd^{\ad_q}\D_{\a_q}\bar{u}_{\a(q-1)\ad(q)}\\
&\hspace{5.5cm}+c_{4}^{(q)}~u^{\a(q)\ad(q-1)}\D_{\a_q}\Dd^{\ad_q}\bar{u}_{\a(q-1)\ad(q)}\Bigl.\Bigr]
\Biggl.\Biggr\}~.
\end{IEEEeqnarray*}
The coefficient of the mass interaction between $\chi^{(q)}$ and $u^{(q)}$ can be set to one by adjusting the
normalization of $u^{(q)}$. Similarly, one of the two $b^{(q+1)}$ coefficients can also be set to one by
adjusting the normalization of $\chi^{(q)}$.
The equations of motion generated by the above action are
\begin{IEEEeqnarray*}{ll}
\E^{(H)}_{\a(s)\ad(s)}=&2\D^{\g}\Dd^2\D_{\g}H_{\a(s)\ad(s)}+2m^2H_{\a(s)\ad(s)}
-\frac{2}{s!}~\Dd_{(\ad_s}\D^2\chi_{\a(s)\ad(s-1))}+\frac{2}{s!}~\D_{(\a_s}\Dd^2\bar{\chi}_{\a(s-1))\ad(s)}\n
\label{EH}\\[2mm]
\E^{(\chi,s)}_{\a(s)\ad(s-1)}=&-2\D^2\Dd^{\ad_s}H_{\a(s)\ad(s)}-2~\frac{s+1}{s}~\D^2\chi_{\a(s)\ad(s-1)}
+\frac{2}{s!}~\D_{(\a_s}\Dd^{\ad_s}\bar{\chi}_{\a(s-1))\ad(s)}+m~u_{\a(s)\ad(s-1)}~~~~\n\label{Exs}\\[2mm]
\E^{(u,s)}_{\a(s)\ad(s-1)}=&2c_1^{(s)}~\Dd^2u_{\a(s)\ad(s-1)}+2c_2^{(s)}~\D^2u_{\a(s)\ad(s-1)}
+\frac{c_3^{(s)}}{s!}~\Dd^{\ad_s}\D_{(\a_s}\bar{u}_{\a(s-1))\ad(s)}\n\label{Eus}\\
&+\frac{c_4^{(s)}}{s!}~\D_{(\a_s}\Dd^{\ad_s}\bar{u}_{\a(s-1))\ad(s)}+m~\chi_{\a(s)\ad(s-1)}
+\frac{b_1^{(s)}}{s!(s-1)!}~\Dd_{(\ad_{s-1}}\D_{(\a_s}\chi_{\a(s-1))\ad(s-2))}\\
&+\frac{b_2^{(s)}}{s!(s-1)!}~\D_{(\a_s}\Dd_{(\ad_{s-1}}\chi_{\a(s-1))\ad(s-2))}
\end{IEEEeqnarray*}
and
\begin{IEEEeqnarray*}{ll}
\E^{(\chi,q)}_{\a(q)\ad(q-1)}=&-b_1^{(q+1)}\D^{\a_{q+1}}\Dd^{\ad_{q}}u_{\a(q+1)\ad(q)}
-b_2^{(q+1)}\Dd^{\ad_{q}}\D^{\a_{q+1}}u_{\a(q+1)\ad(q)}\n\label{Exq}\\
&+2d_1^{(q)}~\Dd^2\chi_{\a(q)\ad(q-1)}+2d_2^{(q)}~\D^2\chi_{\a(q)\ad(q-1)}
+\frac{d_3^{(q)}}{q!}~\Dd^{\ad_q}\D_{(\a_q}\bar{\chi}_{\a(q-1))\ad(q)}\\
&+\frac{d_4^{(q)}}{q!}~\D_{(\a_q}\Dd^{\ad_q}\bar{\chi}_{\a(q-1))\ad(q)}+m u_{\a(q)\ad(q-1)}\\[2mm]
\E^{(u,q)}_{\a(q)\ad(q-1)}=&2c_1^{(q)}~\Dd^2u_{\a(q)\ad(q-1)}+2c_2^{(q)}~\D^2u_{\a(q)\ad(q-1)}
+\frac{c_3^{(q)}}{q!}~\Dd^{\ad_q}\D_{(\a_q}\bar{u}_{\a(q-1))\ad(q)}\n\label{Euq}\\
&+\frac{c_4^{(q)}}{q!}~\D_{(\a_q}\Dd^{\ad_q}\bar{u}_{\a(q-1))\ad(q)}+m~\chi_{\a(q)\ad(q-1)}
+\frac{b_1^{(q)}}{q!(q-1)!}~\Dd_{(\ad_{q-1}}\D_{(\a_q}\chi_{\a(q-1))\ad(q-2))}\\
&+\frac{b_2^{(q)}}{q!(q-1)!}~\D_{(\a_q}\Dd_{(\ad_{q-1}}\chi_{\a(q-1))\ad(q-2))}
\end{IEEEeqnarray*}
for $q=1,2,\dots,s-1$.
Our strategy is to use these equations of motion with an appropriate choice of coefficients
in order to show that on-shell all auxiliary superfields vanish
and we dynamically generate constraints (\ref{H}):
\begin{IEEEeqnarray*}{l}
u_\a=0~\Rightarrow~\chi_\a=0~\Rightarrow~\dots~\Rightarrow~u_{\a(q)\ad(q-1)}=0~\Rightarrow~
\chi_{\a(q)\ad(q-1)}=0~\Rightarrow~\dots\n\label{strategy}\\
\Rightarrow~u_{\a(s)\ad(s-1)}=0~\Rightarrow~\chi_{\a(s)\ad(s-1)}=0~\Rightarrow~\D^{\a_s}H_{\a(s)\ad(s)}=0~,~
\Box H_{\a(s)\ad(s)}=m^2H_{\a(s)\ad(s)}~.
\end{IEEEeqnarray*}
We will do that in steps and recursively. For example, if we assume that $u^{(r)}=\chi^{(r)}=0$ for
$r=1,2,\dots,s$ then we can easily generate (\ref{H}). If $u^{(s)}=\chi^{(s)}=0$
we get via (\ref{Exs}) that on-shell $\D^2\Dd^{\ad_s}H_{\a(s)\ad(s)}=0$. This will lead to
$\D^{\a_s}H_{\a(s)\ad(s)}=0$ based on $\D^{\a_s}\E^{(H)}_{\a(s)\ad(s)}=0$ and
$\E^{(H)}_{\a(s)\ad(s)}=0$ will give $\Box H_{\a(s)\ad(s)}=m^2H_{\a(s)\ad(s)}$.
The next step is to assume that
$u^{(r)}=\chi^{(r)}=0$ for $r=1,2,\dots,s-1$ and show that there is a choice of $c^{(s)}$ coefficients
such that $u^{(s)},~\chi^{(s)}$ both vanish on-shell. We will show that this approach can be iterated
until we reach the bottom of the sequence (\ref{strategy}) and in the process we determine all coefficients.
In every iteration we assume that all auxiliary superfields up to some level can be set to zero and then prove
that there is an appropriate choice of coefficients that will make the next level auxiliary superfields vanish
too.

Because the auxiliary superfields are required to vanish on-shell, one is allowed to consider a linear
redefinition of the auxiliary superfields
$\hat{u}^{(q)}=A u^{(q)}+B \chi^{(q)}$, $\hat{\chi}^{(q)}=C u^{(q)}+D \chi^{(q)}$ without affecting the
outcome. Besides this redefinition being invertible ($AD-CB\neq0$), the coefficients $(A, B, C, D)$ are
arbitrary and can be used to fix any four, q-level coefficients in the action. This freedom has already being
used in the ansatz \eqref{mS}.

A tool that will be used often in this method is the following.
Notice that $\chi^{(q)}$ superfield appears algebraically in $\E^{(u,q)}$. Hence, we can use this to
eliminate all $\chi^{(q)}$ dependence in $\E^{(\chi,q)}$, for all values of $q$. One way of doing that is to
define the following quantity:
\begin{IEEEeqnarray*}{l}
    \I^{(q)}_{\a(q)\ad(q-1)}=m\E^{(\chi,q)}_{\a(q)\ad(q-1)}-2d_1^{(q)}\Dd^2\E^{(u,q)}_{\a(q)\ad(q-1)}
    -2d_2^{(q)}\D^2\E^{u,q)}_{\a(q)\ad(q-1)}\n\label{defIq}\\[1mm]
    \hspace{45mm}-\frac{d_3^{(q)}}{q!}\Dd^{\ad_q}\D_{(\a_q}\bar{\E}^{(u,q)}_{\a(q-1))\ad(q)}
    -\frac{d_4^{(q)}}{q!}\D_{(\a_q}\Dd^{\ad_q}\bar{\E}^{(u,q)}_{\a(q-1))\ad(q)}~.
\end{IEEEeqnarray*}
Plugging in equations (\ref{Exq}) and (\ref{Euq}) we find that
\begin{IEEEeqnarray*}{ll}
    \I^{(q)}_{\a(q)\ad(q-1)}=&
    -m~b_1^{(q+1)}~\D^{\a_{q+1}}\Dd^{\ad_q}u_{\a(q+1)\ad(q)}
    -m~b_2^{(q+1)}~\Dd^{\ad_q}\D^{\a_{q+1}}u_{\a(q+1)\ad(q)}\n\label{Iq}\\[2mm]
    &+m^2~u_{\a(q)\ad(q-1)}+\left\{d_3^{(q)}c_3^{(q)}\right\}~\D^{\g}\Dd^2\D_{\g}u_{\a(q)\ad(q-1)}\\[1mm]
    &+\left\{-4d_1^{(q)}c_2^{(q)}-d_3^{(q)}c_4^{(q)}\right\}~\Dd^2\D^2u_{\a(q)\ad(q-1)}
    +\left\{-4d_2^{(q)}c_1^{(q)}-d_4^{(q)}c_3^{(q)}\right\}~\D^2\Dd^2u_{\a(q)\ad(q-1)}\\[1mm]
    &+\left\{-2d_1^{(q)}c_4^{(q)}-2d_3^{(q)}c_2^{(q)}\right\}
    ~\frac{1}{q!}~\Dd^2\D_{(\a_q}\Dd^{\ad_q}\bar{u}_{\a(q-1))\ad(q)}\\[1mm]
    &+\left\{-2d_2^{(q)}c_3^{(q)}-2d_4^{(q)}c_1^{(q)}\right\}
    ~\frac{1}{q!}~\D^2\Dd^{\ad_q}\D_{(\a_q}\bar{u}_{\a(q-1))\ad(q)}\\[1mm]
    &+\left\{d_3^{(q)}c_4^{(q)}-d_4^{(q)}\left(\frac{q+1}{q}~c_4^{(q)}-c_3^{(q)}\right)\right\}
    ~\frac{1}{q!}~\D_{(\a_q}\Dd^2\D^{\b}u_{\b\a(q-1))\ad(q-1)}\\[1mm]
    &+\left\{\frac{q-1}{q}~d_3^{(q)}c_{3}^{(q)}\right\}
    ~\frac{1}{(q-1)!}~\Dd_{(\ad_{q-1}}\D^2\Dd^{\bd}u_{\a(q)\bd\ad(q-2))}\\[1mm]
    &+\left\{-\frac{q-1}{q}~d_4^{(q)}c_3^{(q)}\right\}
    ~\frac{1}{q!(q-1)!}~\D_{(\a_q}\Dd_{(\ad_{q-1}}\D^{\b}\Dd^{\bd}u_{\b\a(q-1))\bd\ad(q-2))}\\[1mm]
    &+\left\{-\frac{q-1}{q}~d_3^{(q)}c_4^{(q)}\right\}
    ~\frac{1}{q!(q-1)!}~\Dd_{(\ad_{q-1}}\D_{(\a_q}\Dd^{\bd}\D^{\b}u_{\b\a(q-1))\bd\ad(q-2))}\\[2mm]
    &+\left\{-2d_1^{(q)}b_2^{(q)}\right\}
    ~\frac{1}{q!(q-1)!}~\Dd^2\D_{(\a_q}\Dd_{(\ad_{q-1}}\chi_{\a(q-1))\ad(q-2))}\\[1mm]
    &+\left\{-2d_2^{(q)}b_1^{(q)}\right\}
    ~\frac{1}{q!(q-1)!}~\D^2\Dd_{(\ad_{q-1}}\D_{(\a_q}\chi_{\a(q-1))\ad(q-2))}\\[1mm]
    &+\left\{-d_3^{(q)}b_2^{(q)}-d_4^{(q)}\left(b_1^{(q)}-\frac{q+1}{q}~b_2^{(q)}\right)\right\}
    ~\frac{1}{q!}~\D_{(\a_q}\Dd^2\D_{\a_{q-1}}\bar{\chi}_{\a(q-2))\ad(q-1)}\\[1mm]
    &+\left\{\frac{q-1}{q}~d_4^{(q)}b_1^{(q)}\right\}
    ~\frac{1}{q!(q-1)!}~\D_{(\a_q}\Dd_{(\ad_{q-1}}\D_{\a_{q-1}}\Dd^{\gd}\bar{\chi}_{\a(q-2))\gd\ad(q-1))}\\[1mm]
    &+\left\{\frac{q-1}{q}~d_3^{(q)}b_2^{(q)}\right\}
    ~\frac{1}{q!(q-1)!}~\Dd_{(\ad_{q-1}}\D_{(\a_q}\Dd^{\gd}\D_{\a_{q-1}}\bar{\chi}_{\a(q-2))\gd\ad(q-1))}~.
\end{IEEEeqnarray*}
This expression holds for all values of $q=1,2,\dots,s-1$\footnote{Note that for $q=1$, there are no
    $b_1^{(q=1)}$ and $b_2^{(q=1)}$ coupling constants, as can be seen in (\ref{mS}). That is because the
    lowest rank $\chi$ superfield is $\chi^{(1)}$ and there is no $\chi^{(0)}$. Hence for that case the last
    five lines of $(\ref{Iq})|_{q=1}$ drop out. To simplify, things we can keep using the same expression with
the convention the $b_1^{(q=1)}=b_2^{(q=1)}=0$.}. However, the validity of (\ref{defIq}) can be expanded to
include the $q=s$ case as well. In that case, expression \eqref{Iq} remains valid if we assign to the
$d^{(s)}$ constants the corresponding values coming from the massless theory \eqref{masslessS}
\begin{equation}
    d_1^{(s)}=0~,~d_2^{(s)}=-\frac{s+1}{s}~,~d_3^{(s)}=0~,~d_4^{(s)}=2
\end{equation}
and replace the first line with the correct $H_{\a(s)\ad(s)}$ terms
\begin{IEEEeqnarray*}{ll}
    \I^{(s)}_{\a(s)\ad(s-1)}=&
    -2m~\D^2\Dd^{\ad_s}H_{\a(s)\ad(s)}\n\label{Is}\\[2mm]
     &+m^2~u_{\a(s)\ad(s-1)}\\[1mm]
    &+\left\{-4d_2^{(s)}c_1^{(s)}-d_4^{(s)}c_3^{(s)}\right\}~\D^2\Dd^2u_{\a(s)\ad(s-1)}\\[1mm]
    &+\left\{-2d_2^{(s)}c_3^{(s)}-2d_4^{(s)}c_1^{(s)}\right\}
    ~\frac{1}{s!}~\D^2\Dd^{\ad_s}\D_{(\a_s}\bar{u}_{\a(s-1))\ad(s)}\\[1mm]
    &+\left\{-d_4^{(s)}\left(\frac{s+1}{s}~c_4^{(s)}-c_3^{(s)}\right)\right\}
    ~\frac{1}{s!}~\D_{(\a_s}\Dd^2\D^{\b}u_{\b\a(s-1))\ad(s-1)}\\[1mm]
    &+\left\{-\frac{s-1}{s}~d_4^{(s)}c_3^{(s)}\right\}
    ~\frac{1}{s!(s-1)!}~\D_{(\a_s}\Dd_{(\ad_{s-1}}\D^{\b}\Dd^{\bd}u_{\b\a(s-1))\bd\ad(s-2))}\\[2mm]
    &+\left\{-2d_2^{(s)}b_1^{(s)}\right\}
    ~\frac{1}{s!(s-1)!}~\D^2\Dd_{(\ad_{s-1}}\D_{(\a_s}\chi_{\a(s-1))\ad(s-2))}\\[1mm]
    &+\left\{-d_4^{(s)}\left(b_1^{(s)}-\frac{s+1}{s}~b_2^{(s)}\right)\right\}
    ~\frac{1}{s!}~\D_{(\a_s}\Dd^2\D_{\a_{s-1}}\bar{\chi}_{\a(s-2))\ad(s-1)}\\[1mm]
    &+\left\{\frac{s-1}{s}~d_4^{(s)}b_1^{(s)}\right\}
    ~\frac{1}{s!(s-1)!}
    ~\D_{(\a_s}\Dd_{(\ad_{s-1}}\D_{\a_{s-1}}\Dd^{\gd}\bar{\chi}_{\a(s-2))\gd\ad(s-1))}~.
\end{IEEEeqnarray*}
\subsection{Linearized massive supergravity and $s$-th level auxiliary superfields}
\label{subsec:sthlevel}
The term in the first line of \eqref{Is} can also be generated by $\D^2\Dd^{\ad_s}\E^{(H)}_{\a(s)\ad(s)}$
using the mass term of $H_{\a(s)\ad(s)}$
\begin{IEEEeqnarray*}{ll}
    \D^2\Dd^{\ad_s}\E^{(H)}_{\a(s)\ad(s)}=&-2~\D^2\Dd^2\D^2\Dd^{\ad_s}H_{\a(s)\ad(s)}
    +2m^2~\D^2\Dd^{\ad_s}H_{\a(s)\ad(s)}\n\label{D2DdEH}\\[1mm]
    &-2~\frac{s+1}{s}~\D^2\Dd^2\D^2\chi_{\a(s)\ad(s-1)}
    +\frac{2}{s!}~\D^2\Dd^2\D_{(\a_s}\Dd^{\ad_s}\bar{\chi}_{\a(s-1))\ad(s)}~.
\end{IEEEeqnarray*}
Unfortunately, we also generate additional, unwanted $H$ and $\chi^{(s)}$ terms. However, notice
that all these additional terms can be canceled by $\D^2\Dd^2\E^{(\chi,s)}_{\a(s)\ad(s-1)}$.
If we consider the combination
\begin{IEEEeqnarray*}{rl}
\K^{(s)}_{\a(s)\ad(s-1)}=&\D^2\Dd^{\ad(s)}\E^{(H)}_{\a(s)\ad(s)}
+f_1^{(s)}~\D^2\Dd^2\E^{(\chi,s)}_{\a(s)\ad(s-1)}\n\label{preKs}\\[2mm]
=&\left\{-2-2f_1^{(s)}\right\}~\D^2\Dd^2\D^2\Dd^{\ad_s}H_{\a(s)\ad(s)}\\[1mm]
&+\left\{-2\frac{s+1}{s}-2\frac{s+1}{s}f_1^{(s)}\right\}~\D^2\Dd^2\D^2\chi_{\a(s)\ad(s-1)}\\[1mm]
&+\left\{2+2f_1^{(s)}\right\}~\frac{1}{s!}~\D^2\Dd^2\D_{(\a_s}\Dd^{\ad_{s}}\bar{\chi}_{\a(s-1))\ad(s)}\\[1mm]
&+2m^2~\D^2\Dd^{\ad_s}H_{\a(s)\ad(s)}~+~f_1^{(s)}m~\D^2\Dd^2u_{\a(s)\ad(s-1)}~
\end{IEEEeqnarray*}
then obviously the choice $f_1^{(s)}=-1$ is the appropriate one because it cancels all the unwanted terms
and we are left with terms that are proportional to $m$ and $m^2$.
These cancellations are not an accident but a consequence of
the gauge invariance of the massless action \eqref{masslessS}. Because of \eqref{dH} and \eqref{dx}
the massless theory equations of motion must satisfy the following Jacobi identity
\begin{equation}
    \Dd^{\ad_s}\E^{(H,m=0)}_{\a(s)\ad(s)}=\Dd^2\E^{(\chi,s,m=0)}_{\a(s)\ad(s-1)}
\end{equation}
hence, the specific combination
$\K^{(s)}_{\a(s)\ad(s-1)}=\D^2\Dd^{\ad(s)}\E^{(H)}_{\a(s)\ad(s)}-\D^2\Dd^2\E^{(\chi,s)}_{\a(s)\ad(s-1)}$
must include only terms proportional to $m$ and $m^2$ which identically vanish in the massless limit
\begin{IEEEeqnarray*}{l}
\K^{(s)}_{\a(s)\ad(s-1)}=2m^2~\D^2\Dd^{\ad_s}H_{\a(s)\ad(s)}~-m~\D^2\Dd^2u_{\a(s)\ad(s-1)}~.\n\label{Ks}
\end{IEEEeqnarray*}
Now we can combining \eqref{Is} with \eqref{Ks} in order to cancel the common $H$-terms
\begin{IEEEeqnarray*}{rl}
    \J^{(s)}_{\a(s)\ad(s-1)}=&\frac{1}{m}~\K^{(s)}_{\a(s)\ad(s-1)}+\I^{(s)}_{\a(s)\ad(s-1)}\n\label{Js}\\[2mm]
    =&m^2~u_{\a(s)\ad(s-1)}\\[2mm]
    &+\left\{-4d_2^{(s)}c_1^{(s)}-d_4^{(s)}c_3^{(s)}-1\right\}~\D^2\Dd^2u_{\a(s)\ad(s-1)}\\[1mm]
    &+\left\{-2d_2^{(s)}c_3^{(s)}-2d_4^{(s)}c_1^{(s)}\right\}
    ~\frac{1}{s!}~\D^2\Dd^{\ad_s}\D_{(\a_s}\bar{u}_{\a(s-1))\ad(s)}\\[1mm]
    &+\left\{-d_4^{(s)}\left(\frac{s+1}{s}~c_4^{(s)}-c_3^{(s)}\right)\right\}
    ~\frac{1}{s!}~\D_{(\a_s}\Dd^2\D^{\b}u_{\b\a(s-1))\ad(s-1)}\\[1mm]
    &+\left\{-\frac{s-1}{s}~d_4^{(s)}c_3^{(s)}\right\}
    ~\frac{1}{s!(s-1)!}~\D_{(\a_s}\Dd_{(\ad_{s-1}}\D^{\b}\Dd^{\bd}u_{\b\a(s-1))\bd\ad(s-2))}\\[2mm]
    &+\left\{-2d_2^{(s)}b_1^{(s)}\right\}
    ~\frac{1}{s!(s-1)!}~\D^2\Dd_{(\ad_{s-1}}\D_{(\a_s}\chi_{\a(s-1))\ad(s-2))}\\[1mm]
    &+\left\{-d_4^{(s)}\left(b_1^{(s)}-\frac{s+1}{s}~b_2^{(s)}\right)\right\}
    ~\frac{1}{s!}~\D_{(\a_s}\Dd^2\D_{\a_{s-1}}\bar{\chi}_{\a(s-2))\ad(s-1)}\\[1mm]
    &+\left\{\frac{s-1}{s}~d_4^{(s)}b_1^{(s)}\right\}
    ~\frac{1}{s!(s-1)!}~\D_{(\a_s}\Dd_{(\ad_{s-1}}\D_{\a_{s-1}}\Dd^{\gd}\bar{\chi}_{\a(s-2))\gd\ad(s-1))}~.
\end{IEEEeqnarray*}
This gives an equation that depends only on $u^{(s)}$ and $\chi^{(s-1)}$ and includes an algebraic term for
$u^{(s)}$. On-shell the left hand side of the
equation vanishes, because $\J^{(s)}$  is constructed by equations of motions.
Therefore by tuning the various coefficients we can use the algebraic term in order to make
$u^{(s)}$ vanish.

For the special case of $s=1$ (linearized supergravity), the last three lines drop out and the fourth one
identically vanish
\begin{IEEEeqnarray*}{rl}\n
    \J^{(1)}_{\a}
    =m^2~u_{\a}~
    &+\left\{-4d_2^{(1)}c_1^{(1)}-d_4^{(1)}c_3^{(1)}-1\right\}~\D^2\Dd^2u_{\a}\\[1mm]
    &+\left\{-2d_2^{(1)}c_3^{(1)}-2d_4^{(1)}c_1^{(1)}\right\}
    ~\D^2\Dd^{\ad}\D_{\a}\bar{u}_{\ad}\\[1mm]
    &+\left\{-d_4^{(1)}\left(2~c_4^{(1)}-c_3^{(1)}\right)\right\}
    ~\D_{\a}\Dd^2\D^{\b}u_{\b}~.
\end{IEEEeqnarray*}
Therefore, in order to have $u_{\a}$ vanishing on-shell we must select the coefficients to satisfy
\begin{equation}
    \begin{rcases*}
4~d_2^{(1)}c_1^{(1)}+d_4^{(1)}c_3^{(1)}=-1\\
d_2^{(1)}c_3^{(1)}+d_4^{(1)}c_1^{(1)}=0\\
2~c_4^{(1)}-c_3^{(1)}=0
\end{rcases*}\Rightarrow~c_1^{(1)}=\frac{1}{6}~,~c_3^{(1)}=\frac{1}{6}~,~c_4^{(1)}=\frac{1}{12}~.
\end{equation}
These are precisely the results found in~\cite{Gates:2013tka}.

For the general case (arbitrary $s$), we assume that all auxiliary superfields of lower levels vanish on-shell
$u^{(r)}=\chi^{(r)}=0$ for $r=1,2,\dots,s-1$. Then because of $\E^{(\chi,s-1)}_{\a(s-1)\ad(s-2)}=0$,
superfield $u_{\a(s)\ad(s-1)}$ will satisfy the following constraint
\begin{equation}
b_1^{(s)}~\D^{\a_s}\Dd^{\ad_{s-1}}u_{\a(s)\ad(s-1)}
+b_2^{(s)}~\Dd^{\ad_{s-1}}\D^{\a_s}u_{\a(s)\ad(s-1)}=0\label{con-s}
\end{equation}
which leads to the identification of various terms in \eqref{Js}
\begin{IEEEeqnarray*}{rl}
    \J^{(s)}_{\a(s)\ad(s-1)}=
    &m^2~u_{\a(s)\ad(s-1)}
   ~+\left\{-4d_2^{(s)}c_1^{(s)}-d_4^{(s)}c_3^{(s)}-1\right\}~\D^2\Dd^2u_{\a(s)\ad(s-1)}\n\label{Jssimple}
    \\[1mm]
    &+\left\{-2d_2^{(s)}c_3^{(s)}-2d_4^{(s)}c_1^{(s)}\right\}
    ~\frac{1}{s!}~\D^2\Dd^{\ad_s}\D_{(\a_s}\bar{u}_{\a(s-1))\ad(s)}\\[1mm]
&+\left\{-d_4^{(s)}\left(\frac{s+1}{s}~c_4^{(s)}-c_3^{(s)}\right)
-\frac{s-1}{s}~d_4^{(s)}c_3^{(s)}\frac{b_2^{(s)}}{b_1^{(s)}}\right\}
    ~\frac{1}{s!}~\D_{(\a_s}\Dd^2\D^{\b}u_{\b\a(s-1))\ad(s-1)}~.
\end{IEEEeqnarray*}
Therefore, in order to be able to make $u^{(s)}$ vanish on-shell we must choose the coefficients such that
\begin{equation}\label{c(s)}
    \begin{rcases*}
        4~d_2^{(s)}~c_1^{(s)}+d_4^{(s)}~c_3^{(s)}=f_1^{(s)}=-1\\
        d_2^{(s)}~c_3^{(s)}+d_4^{(s)}~c_1^{(s)}=0\\
        d_4^{(s)}\left(\frac{s+1}{s}~c_4^{(s)}-c_3^{(s)}\right)
        +\frac{s-1}{s}~d_4^{(s)}~c_3^{(s)}~\frac{b_2^{(s)}}{b_1^{(s)}}=0
\end{rcases*}\Rightarrow
\begin{cases*}
    c_1^{(s)}=\frac{1}{4}~\frac{s(s+1)}{2s+1}\\
    c_3^{(s)}=\frac{1}{2}~\frac{s^2}{2s+1}\\
    c_4^{(s)}=\frac{1}{2}~\frac{s^3}{(s+1)(2s+1)}~\left(1-\frac{s-1}{s}~\frac{b_2^{(s)}}{b_1^{(s)}}\right)
\end{cases*}
\end{equation}
Notice that as an input in the above system of equations are the $d^{(s)}$ and $f_1^{(s)}$ coefficients fixed
by the massless action and Jacobi identity. This will be a repeating pattern that will become explicit
in section (\ref{subsec:qthlevel}). Moreover, $c_2^{(s)}$ remains arbitrary and irrelevant to $u^{(s)}=0$
which is consistent with the findings of~\cite{Gates:2013tka}, thus we can set it to zero.
Also notice that the coefficients $b_1^{(s)},~b_2^{(s)}$ have not being
fixed yet, however they participate in the combination $b_2^{(s)}/b_1^{(s)}$
which demands $b_1^{(s)}$ to be non-zero. Hence, using the freedom to normalize $\chi^{(s-1)}$ accordingly we
can set it to one, $b_1^{(s)}=1$. Finally, the vanishing of $\chi^{(s)}$ on-shell follows automatically
from equation $\E^{(u,s)}=0$.
\subsection{$\Ysf=5/2$ supermultiplet and $(s-1)$-th level auxiliary superfields}
\label{subsec:(s-1)thlevel}
Now we repeat the process for $(s-1)$-th level auxiliary superfields.
The main goal is to make $u^{(s-1)}$ vanish
on-shell. If it does then, $\chi^{(s-1)}$ will automatically follow as demonstrated above.
We start with (\ref{Js}) and use it to calculate $\D^{\a_s}\Dd^{\ad_{s-1}}\J^{(s)}_{\a(s)\ad(s-1)}$
and $\Dd^{\ad_{s-1}}\D^{\a_s}\J^{(s)}_{\a(s)\ad(s-1)}$ in order to generate the
$\D^{\a_s}\Dd^{\ad_{s-1}}u_{\a(s)\ad(s-1)}$, $\Dd^{\ad_{s-1}}\D^{\a_s}u_{\a(s)\ad(s-1)}$ terms that appear in
$I^{(s-1)}_{\a(s-1)\ad(s-2)}$\footnote{That is equation (\ref{Iq}) for $q=s-1$.}.
After some algebra one can find that
\begin{IEEEeqnarray*}{rl}
\D^{\a_s}\Dd^{\ad_{s-1}}&\J^{(s)}_{\a(s)\ad(s-1)}+b_2^{(s)}~\Dd^{\ad_{s-1}}\D^{\a_s}\J^{(s)}_{\a(s)\ad(s-1)}
=\n\label{DDdJs}\\[4mm]
=&m^2~\D^{\a_s}\Dd^{\ad_{s-1}}u_{\a(s)\ad(s-1)}+m^2~b_2^{(s)}~\Dd^{\ad_{s-1}}\D^{\a_s}u_{\a(s)\ad(s-1)}\\[1mm]
&+\left\{\frac{2s-1}{s^2}~d_4^{(s)}c_3^{(s)}\right\}~\D^2\Dd^2\D^{\b}\Dd^{\bd}u_{\b\a(s-1)\bd\ad(s-2)}\\[1mm]
&+\left(1-\frac{s+1}{s}~b_2^{(s)}\right)\left\{-\frac{s-1}{s}~d_4^{(s)}c_3^{(s)}\right\}
~\D^{\b}\Dd^2\D^2\Dd^{\bd}u_{\b\a(s-1)\bd\ad(s-1)}\\[1mm]
&+\left(1-\frac{s+1}{s}~b_2^{(s)}\right)\left\{-\frac{s-1}{s}~d_4^{(s)}c_3^{(s)}b_2^{(s)}\right\}
~\Dd^{\bd}\D^2\Dd^2\D^{\b}u_{\b\a(s-1)\bd\ad(s-1)}\\[1mm]
&+\left\{\frac{(s-1)^2}{s^2}~d_4^{(s)}c_3^{(s)}b_2^{(s)}\right\}
~\frac{1}{(s-1)!}~\D_{(\a_{s-1}}\Dd^2\D^{\b}\Dd^{\gd}\D^{\g}u_{\b\g\a(s-2))\gd\ad(s-2)}\\[1mm]
&+\left(1-\frac{s+1}{s}~b_2^{(s)}\right)\left\{\frac{s-2}{s}~d_4^{(s)}c_3^{(s)}\right\}
~\frac{1}{(s-2)!}~\Dd_{(\ad_{s-2}}\D^2\Dd^{\bd}\D^{\g}\Dd^{\gd}u_{\g\a(s-1)\bd\gd\ad(s-3))}\\[1mm]
&+\left\{-\frac{(s-2)(s-1)}{s^2}~d_4^{(s)}c_3^{(s)}\right\}
~\frac{1}{(s-1)!(s-2)!}
~\D_{(\a_{s-1}}\Dd_{(\ad_{s-2}}\D^{\b}\Dd^{\bd}\D^{\g}\Dd^{\gd}u_{\b\g\a(s-2))\bd\gd\ad(s-3))}\\[2mm]
&+\left\{2~\frac{2s-1}{s(s-1)}~d_2^{(s)}\right\}~\D^2\Dd^2\D^2\chi_{\a(s-1)\ad(s-2)}\\[1mm]
&+\left\{-\frac{2s-1}{s^2}~d_4^{(s)}\right\}
~\frac{1}{(s-1)!}~\D^2\Dd^2\D_{(\a_{s-1}}\Dd^{\bd}\bar{\chi}_{\a(s-2))\bd\ad(s-2)}\\[1mm]
&+\left(1-\frac{s+1}{s}~b_2^{(s)}\right)^2d_4^{(s)}~\frac{1}{(s-1)!}
~\Dd^{\bd}\D^2\Dd^2\D_{(\a_{s-1}}\bar{\chi}_{\a(s-2))\bd\ad(s-2)}\\[1mm]
&+\left(2~\frac{(s-1)^2}{s^2}~d_4^{(s)}\right)\left\{1-\frac{s+1}{s-1}~b_2^{(s)}\right\}
~\frac{1}{(s-1)!}~\D_{(\a_{s-1}}\Dd^2\D^2\Dd^{\bd}\bar{\chi}_{\a(s-2))\bd\ad(s-2)}\\[1mm]
&+\left(1-\frac{s+1}{s}~b_2^{(s)}\right)\left\{-\frac{s-2}{s}~d_4^{(s)}\right\}
~\frac{1}{(s-1)!(s-2)!}
~\Dd_{(\ad_{s-2}}\D^2\Dd^{\bd}\D_{(\a_{s-1}}\Dd^{\gd}\bar{\chi}_{\a(s-2))\bd\gd\ad(s-3))}\\[1mm]
&+\left(1-\frac{s+1}{s}~b_2^{(s)}\right)\left\{-\frac{s-2}{s}~d_4^{(s)}\right\}
~\frac{1}{(s-1)!}
~\D_{(\a_{s-1}}\Dd^2\D_{\a_{s-2}}\Dd^{\bd}\D^{\b}\bar{\chi}_{\b\a(s-3)))\bd\ad(s-2)}\\[1mm]
&+\left\{-2~\frac{s-2}{s}~d_2^{(s)}\right\}
~\frac{1}{(s-1)!(s-2)!}
~\D_{(\a_{s-1}}\Dd_{(\ad_{s-2}}\D^{\b}\Dd^{\bd}\D^2\chi_{\b\a(s-2))\bd\ad(s-3))}\\[1mm]
&+\left\{\frac{(s-2)^2}{s^2}~d_4^{(s)}\right\}
~\frac{1}{(s-1)!(s-2)!}
~\D_{(\a_{s-1}}\Dd_{(\ad_{s-2}}\D_{\a_{s-2}}\Dd^{\bd}\D^{\g}\Dd^{\gd}\bar{\chi}_{\g\a(s-3))\bd\gd\ad(s-3))}~.
\end{IEEEeqnarray*}
Once again, besides the desired terms of the first line, we generated many more $u^{(s)}$ and $\chi^{(s-1)}$
terms. However, similarly to (\ref{D2DdEH}) and (\ref{preKs}), all these terms can be absorbed by
appropriate use of $\E^{(\chi,s-1)}$ and leave only terms that depend on the mass parameter. A careful examination of
the terms in (\ref{DDdJs}) suggests that we have to consider the following
\begin{IEEEeqnarray*}{rl}
    \K^{(s-1)}_{\a(s-1)\ad(s-2)}&=
\D^{\a_s}\Dd^{\ad_{s-1}}\J^{(s)}_{\a(s)\ad(s-1)}+b_2^{(s)}~\Dd^{\ad_{s-1}}\D^{\a_s}\J^{(s)}_{\a(s)\ad(s-1)}
+f_1^{(s-1)}~\D^2\Dd^2\E^{(\chi,s-1)}_{\a(s-1)\ad(s-2)}\n\\
+&f_2^{(s-1)}~\D^{\g}\Dd^2\D_{\g}\E^{(\chi,s-1)}_{\a(s-1)\ad(s-2)}
+\frac{f_3^{(s-1)}}{(s-2)!}~\Dd_{(\ad_{s-2}}\D^2\Dd^{\bd}\E^{(\chi,s-1)}_{\a(s-1)\bd\ad(s-3))}\\
+&\frac{f_4^{(s-1)}}{(s-1)!}~\D_{(\a_{s-1}}\Dd^2\D^{\b}\E^{(\chi,s-1)}_{\b\a(s-2))\ad(s-2)}
+\frac{f_5^{(s-1)}}{(s-1)!(s-2)!}~\D_{(\a_{s-1}}\Dd_{(\ad_{s-2}}\D^{\b}\Dd^{\bd}
\E^{(\chi,s-1)}_{\b\a(s-2))\bd\ad(s-3))}
\end{IEEEeqnarray*}
for appropriate values of $f_1^{(s-1)},~f_2^{(s-1)},~f_3^{(s-1)},~f_4^{(s-1)},~f_5^{(s-1)}$.
Notice, that many terms in (\ref{DDdJs}) cancel if we tune $b_2^{(s)}$ such that
\begin{equation}
    1-\frac{s+1}{s}~b_2^{(s)}=0~\Rightarrow~b_2^{(s)}=\frac{s}{s+1}~.\label{b2}
\end{equation}
With this choice, the contributions proportional to $f_2^{(s-1)},~f_3^{(s-1)}$ are not
relevant anymore and will be ignored ($f_2^{(s-1)}=f_3^{(s-1)}=0$)
\begin{IEEEeqnarray*}{l}
    \K^{(s-1)}_{\a(s-1)\ad(s-2)}=\n\label{K(s-1)}\\[3mm]
=m^2~\D^{\a_s}\Dd^{\ad_{s-1}}u_{\a(s)\ad(s-1)}+m^2~b_2^{(s)}~\Dd^{\ad_{s-1}}\D^{\a_s}u_{\a(s)\ad(s-1)}\\[1mm]
+m f_1^{(s-1)}\D^2\Dd^2u_{\a(s-1)\ad(s-2)}
+m~\frac{f_4^{(s-1)}}{(s-1)!}~\D_{(\a_{s-1}}\Dd^2\D^{\b}u_{\b\a(s-2))\ad(s-2)}\\[1mm]
+m~\frac{f_5^{(s-1)}}{(s-1)!(s-2)!}
~\D_{(\a_{s-1}}\Dd_{(\ad_{s-2}}\D^{\b}\Dd^{\bd}u_{\b\a(s-2))\bd\ad(s-3))}\\[3mm]
+\left\{\frac{2s-1}{s^2}~d_4^{(s)}c_3^{(s)}-f_1^{(s-1)}\right\}
~\D^2\Dd^2\D^{\b}\Dd^{\bd}u_{\b\a(s-1)\bd\ad(s-2)}\\[1mm]
+\left\{\frac{(s-1)^2}{s^2}~d_4^{(s)}c_3^{(s)}b_2^{(s)}-f_4^{(s-1)}b_2^{(s)}\right\}
~\frac{1}{(s-1)!}~\D_{(\a_{s-1}}\Dd^2\D^{\b}\Dd^{\gd}\D^{\g}u_{\b\g\a(s-2))\gd\ad(s-2)}\\[1mm]
+\left\{-\frac{(s-2)(s-1)}{s^2}~d_4^{(s)}c_3^{(s)}-f_5^{(s-1)}\right\}
~\frac{1}{(s-1)!(s-2)!}
~\D_{(\a_{s-1}}\Dd_{(\ad_{s-2}}\D^{\b}\Dd^{\bd}\D^{\g}\Dd^{\gd}u_{\b\g\a(s-2))\bd\gd\ad(s-3))}\\[3mm]
+\left\{2~\frac{2s-1}{s(s-1)}~d_2^{(s)}+2f_1^{(s-1)}d_2^{(s-1)}\right\}
~\D^2\Dd^2\D^2\chi_{\a(s-1)\ad(s-2)}\\[1mm]
+\left\{-\frac{2s-1}{s^2}~d_4^{(s)}+f_1^{(s-1)}d_4^{(s-1)}\right\}
~\frac{1}{(s-1)!}~\D^2\Dd^2\D_{(\a_{s-1}}\Dd^{\bd}\bar{\chi}_{\a(s-2))\bd\ad(s-2)}\\[1mm]
+\left\{\frac{(s-1)(s-2)}{s^2}~d_4^{(s)}-\frac{(s+1)(s-1)}{s^2}~d_4^{(s)}b_2^{(s)}
+f_4^{(s-1)}\left(\frac{s}{s-1}~d_4^{(s-1)}-d_3^{(s-1)}\right)+f_5^{(s-1)}d_4^{(s-1)}\right\}\times\\
\hfill\times~\frac{1}{(s-1)!}~\D_{(\a_{s-1}}\Dd^2\D^2\Dd^{\bd}\bar{\chi}_{\a(s-2))\bd\ad(s-2)}\\[1mm]
+\left\{\frac{s-2}{s-1}~f_4^{(s-1)}d_3^{(s-1)}\right\}
~\frac{1}{(s-1)!}
~\D_{(\a_{s-1}}\Dd^2\D_{\a_{s-2}}\Dd^{\bd}\D^{\b}\bar{\chi}_{\b\a(s-3)))\bd\ad(s-2)}\\[1mm]
+\left\{-2~\frac{s-2}{s}~d_2^{(s)}+2f_5^{(s-1)}d_2^{(s-1)}\right\}
~\frac{1}{(s-1)!(s-2)!}
~\D_{(\a_{s-1}}\Dd_{(\ad_{s-2}}\D^{\b}\Dd^{\bd}\D^2\chi_{\b\a(s-2))\bd\ad(s-3))}\\[1mm]
+\left\{\frac{(s-2)^2}{s^2}~d_4^{(s)}+\frac{s-2}{s-1}~f_5^{(s-1)}d_4^{(s-1)}\right\}
~\frac{1}{(s-1)!(s-2)!}
~\D_{(\a_{s-1}}\Dd_{(\ad_{s-2}}\D_{\a_{s-2}}\Dd^{\bd}\D^{\g}\Dd^{\gd}\bar{\chi}_{\g\a(s-3))\bd\gd\ad(s-3))}
\end{IEEEeqnarray*}
In order to keep only the terms that depend on mass, we must select the three
$f^{(s-1)}$ in the following way
\begin{IEEEeqnarray*}{l}
    f_1^{(s-1)}=\frac{2s-1}{s^2}~d_4^{(s)}~c_3^{(s)}\n\\[2mm]
    f_4^{(s-1)}=\frac{(s-1)^2}{s^2}~d_4^{(s)}~c_3^{(s)}\n\\[2mm]
    f_5^{(s-1)}=-~\frac{(s-1)(s-2)}{s^2}~d_4^{(s)}~c_3^{(s)}\n
\end{IEEEeqnarray*}
and the $d^{(s-1)}$ parameters must satisfy
\begin{equation}\label{pred(s-1)}
    \begin{rcases*}
        f_1^{(s-1)}~d_2^{(s-1)}=-~\frac{2s-1}{s(s-1)}~d_2^{(s)}\\
        f_5^{(s-1)}~d_2^{(s-1)}=\frac{s-2}{s}~d_2^{(s)}\\[4mm]
        f_1^{(s-1)}~d_4^{(s-1)}=\frac{2s-1}{s^2}~d_4^{(s)}\\
        f_5^{(s-1)}~d_4^{(s-1)}=-~\frac{(s-1)(s-2)}{s^2}~d_4^{(s)}\\[4mm]
        \frac{(s-1)(s-2)}{s^2}~d_4^{(s)}-\frac{(s+1)(s-1)}{s^2}~d_4^{(s)}b_2^{(s)}
        +f_4^{(s-1)}\left(\frac{s}{s-1}~d_4^{(s-1)}-d_3^{(s-1)}\right)+f_5^{(s-1)}d_4^{(s-1)}=0\\
        d_3^{(s)}=0
    \end{rcases*}
\end{equation}
Equations (\ref{pred(s-1)}) are all compatible with each other and have as a solution
\begin{equation}\label{d(s-1)}
        d_2^{(s-1)}=-~\frac{s}{s-1}~\frac{d_2^{(s)}}{d_4^{(s)}~c_3^{(s)}}~,~
        d_3^{(s-1)}=0~,~
        d_4^{(s-1)}=\frac{1}{c_3^{(s)}}~.
\end{equation}
The last step is to combine $\K^{(s-1)}_{\a(s-1)\ad(s-2)}$
with $\I^{(s-1)}_{\a(s-1)\ad(s-2)}$ in order to cancel the $u^{(s)}$ term and construct
$\J^{(s-1)}_{\a(s-1)\ad(s-2)}=\frac{1}{m}~\K^{(s-1)}_{\a(s-1)\ad(s-2)}+\I^{(s-1)}_{\a(s-1)\ad(s-2)}$
\begin{IEEEeqnarray*}{l}
\J^{(s-1)}_{\a(s-1)\ad(s-2)}=m^2~u_{\a(s-1)\ad(s-2)}
+\left\{-4d_2^{(s-1)}c_1^{(s-1)}-d_4^{(s-1)}c_3^{(s-1)}+f_1^{(s-1)}\right\}~\D^2\Dd^2u_{\a(s-1)\ad(s-2)}
\n\label{preJ(s-1)}\\[2mm]
\hspace{1.3cm}+\left\{-4d_1^{(s-1)}c_2^{(s-1)}\right\}\Dd^2\D^2u_{\a(s-1)\ad(s-2)}
+\left\{-2d_1^{(s-1)}c_4^{(s-1)}\right\}
~\frac{1}{(s-1)!}~\Dd^2\D_{(\a_{s-1}}\Dd^{\bd}\bar{u}_{\a(s-2))\bd\ad(s-2)}\\[1mm]
\hspace{1.3cm}+\left\{-2d_2^{(s-1)}c_3^{(s-1)}-2d_4^{(s-1)}c_1^{(s-1)}\right\}
~\frac{1}{(s-1)!}~\D^2\Dd^{\ad_{s-1}}\D_{(\a_{s-1}}\bar{u}_{\a(s-2))\ad(s-1)}\\[1mm]
\hspace{1.3cm}+\left\{-d_4^{(s-1)}\left(\frac{s}{s-1}~c_4^{(s-1)}-c_3^{(s-1)}\right)+f_4^{(s-1)}\right\}
~\frac{1}{(s-1)!}~\D_{(\a_{s-1}}\Dd^2\D^{\b}u_{\b\a(s-2))\ad(s-2)}\\[1mm]
\hspace{1.3cm}+\left\{-\frac{s-2}{s-1}~d_4^{(s-1)}c_3^{(s-1)}+f_5^{(s-1)}\right\}
~\frac{1}{(s-1)!(s-2)!}~\D_{(\a_{s-1}}\Dd_{(\ad_{s-2}}\D^{\b}\Dd^{\bd}u_{\b\a(s-2))\bd\ad(s-3))}\\[1mm]
\hspace{1.3cm}+\left\{-2d_2^{(s-1)}b_1^{(s-1)}\right\}
~\frac{1}{(s-1)!(s-2)!}~\D^2\Dd_{(\ad_{s-2}}\D_{(\a_{s-1}}\chi_{\a(s-2))\ad(s-3))}\\[1mm]
\hspace{1.3cm}+\left\{-2d_1^{(s-1)}b_2^{(s-1)}\right\}
~\frac{1}{(s-1)!(s-2)!}~\Dd^2\D_{(\a_{s-1}}\Dd_{(\ad_{s-2}}\chi_{\a(s-2))\ad(s-3))}\\[1mm]
\hspace{1.3cm}+\left\{-d_4^{(s-1)}\left(b_1^{(s-1)}-\frac{s}{s-1}~b_2^{(s-1)}\right)\right\}
~\frac{1}{(s-1)!}~\D_{(\a_{s-1}}\Dd^2\D_{\a_{s-2}}\bar{\chi}_{\a(s-3))\ad(s-2)}\\[1mm]
\hspace{1.3cm}+\left\{\frac{s-2}{s-1}~d_4^{(s-1)}b_1^{(s-1)}\right\}
~\frac{1}{(s-1)!(s-2)!}
~\D_{(\a_{s-1}}\Dd_{(\ad_{s-2}}\D_{\a_{s-2}}\Dd^{\gd}\bar{\chi}_{\a(s-3))\gd\ad(s-2))}~.
\end{IEEEeqnarray*}
This equation will determine all the $c^{(s-1)}$ coefficients in order to have $u^{(s-1)}$ vanishing on-shell.

For $s=2$, $\Ysf=5/2$ supermultiplet, the last four lines drop out and we are left with three
equations fixing the three unknowns. For the general case, we work under the assumption that we have already set
to zero all lower level auxiliary superfields ($u^{(r)}=\chi^{(r)}=0$  for $r=1,2,\dots,s-2$). Therefore,
using $\E^{(\chi,s-2)}$ we find that $u^{(s-1)}$ on-shell will satisfy  the constraint
\begin{equation}
    b_1^{(s-1)}~\D^{\a_{s-1}}\Dd^{\ad_{s-2}}u_{\a(s-1)\ad(s-2)}
    +b_2^{(s-1)}~\Dd^{\ad_{s-2}}\D^{\a_{s-1}}u_{\a(s-1)\ad(s-2)}=0
\end{equation}
forcing the collapse of various terms in (\ref{preJ(s-1)})
\begin{IEEEeqnarray*}{l}
    \J^{(s-1)}_{\a(s-1)\ad(s-2)}= m^2~u_{\a(s-1)\ad(s-2)}\n\label{J(s-2)}
+\left\{-4d_2^{(s-1)}c_1^{(s-1)}-d_4^{(s-1)}c_3^{(s-1)}+f_1^{(s-1)}\right\}~\D^2\Dd^2u_{\a(q)\ad(q-1)}\\[1mm]
\hspace{1.3cm}+\left\{-4d_1^{(s-1)}c_2^{(s-1)}\right\}\Dd^2\D^2u_{\a(s-1)\ad(s-2)}
+\left\{-2d_1^{(s-1)}c_4^{(s-1)}\right\}
~\frac{1}{(s-1)!}~\Dd^2\D_{(\a_{s-1}}\Dd^{\bd}\bar{u}_{\a(s-2))\bd\ad(s-2)}\\[1mm]
\hspace{1.3cm}+\left\{-2d_2^{(s-1)}c_3^{(s-1)}-2d_4^{(s-1)}c_1^{(s-1)}\right\}
~\frac{1}{(s-1)!}~\D^2\Dd^{\ad_{s-1}}\D_{(\a_{s-1}}\bar{u}_{\a(s-2))\ad(s-1)}\\[1mm]
\hspace{1.3cm}+\left\{-d_4^{(s-1)}\left(\frac{s}{s-1}~c_4^{(s-1)}-c_3^{(s-1)}\right)+f_4^{(s-1)}
+\frac{b_2^{(s-1)}}{b_1^{(s-1)}}~\left(-\frac{s-2}{s-1}~d_4^{(s-1)}c_3^{(s-1)}+f_5^{(s-1)}\right)\right\}\times
\\
\hfill\times~\frac{1}{(s-1)!}~\D_{(\a_{s-1}}\Dd^2\D^{\b}u_{\b\a(s-2))\ad(s-2)}~.
\end{IEEEeqnarray*}
Hence we conclude that we must select coefficients $c_1^{(s-1)},~c_3^{(s-1)},~c_4^{(s-1)}$ as follows
\begin{IEEEeqnarray*}{l}\n
    4~d_2^{(s-1)}~c_1^{(s-1)}+d_4^{(s-1)}~c_3^{(s-1)}=f_1^{(s-1)}~,\sn\\[2mm]
    d_2^{(s-1)}~c_3^{(s-1)}+d_4^{(s-1)}~c_1^{(s-1)}=0~,\sn\\[2mm]
    d_4^{(s-1)}\left(\frac{s}{s-1}~c_4^{(s-1)}-c_3^{(s-1)}\right)
    +\frac{b_2^{(s-1)}}{b_1^{(s-1)}}~\frac{s-2}{s-1}~d_4^{(s-1)}c_3^{(s-1)}
    =f_4^{(s-1)}+\frac{b_2^{(s-1)}}{b_1^{(s-1)}}~f_5^{(s-1)}~,\sn\\[2mm]
    d_1^{(s-1)}~c_2^{(s-1)}=0~,\sn\label{c2(s-1)}\\[2mm]
    d_1^{(s-1)}~c_4^{(s-1)}=0~.\sn
\end{IEEEeqnarray*}
The solution of the above is
\begin{IEEEeqnarray*}{ll}\n\label{c(s-1)}
c_1^{(s-1)}=\frac{1}{16}~\frac{(s+1)s(s-1)(2s-1)}{(2s+1)^2}~,&~
c_3^{(s-1)}=-~\frac{1}{8}~\frac{s(s-1)^2(2s-1)}{(2s+1)^2}~,\sn\\
c_4^{(s-1)}=\frac{1}{8}~\frac{(s-1)^3}{2s+1}\left(1-\frac{s-2}{s-1}~\frac{b_2^{(s-1)}}{b_1^{(s-1)}}\right)~,&~
d_1^{(s-1)}=0~,~c_2^{(s-1)}=\text{arbitrary}~.\sn
\end{IEEEeqnarray*}
Equations (\ref{d(s-1)}) and (\ref{c(s-1)}) fix all the $(s-1)$-level coefficients. Notice that $c_2^{(s-1)}$
remains arbitrary and not relevant to the on-shell vanishing of $u^{(s-1)}$. Similarly to
(\ref{subsec:sthlevel}) we adopt the convention of setting it to zero $(c_{2}^{(s-1)}=0$). However, in this
case equation (\ref{c2(s-1)}) offers an explanation for this freedom. Again the input for the determination
of $c^{(s-1)}$ are the $d^{(s-1)}$ and $f^{(s-1)}$ parameters. The conclusion is that the kinetic energy terms
of superfield $\chi^{(s-1)}$ in (\ref{mS}) have the same structure as the kinetic energy terms of
$\chi^{(s)}$ which are dictated by the massless limit and gauge redundancy. This has the effect of
matching the kinetic energy terms of $u^{(s-1)}$ with those of $u^{(s)}$.
Also, once again, coefficients $b_2^{(s-1)}$ and $b_1^{(s-1)}$ appear only in the combination
$b_2^{(s-1)}/b_1^{(s-1)}$, which demands $b_1^{(s-1)}\neq0$. By using the normalization of
$\chi^{(s-2)}$ we can set it to one, $b_1^{(s-1)}=1$. Finally, the vanishing of $\chi^{(s-1)}$ trivially
follows from equation $\E^{(u,s-1)}=0$.
\subsection{Vanishing of $q$-th level auxiliary superfields}\label{subsec:qthlevel}
The above procedure can be iterated and step by step fix all coefficients such that all auxiliary
superfields can be set to zero on-shell. Briefly, for the $(s-2)$ level we start with
(\ref{preJ(s-1)}) and use it to find $\K^{(s-2)}$ which will include only $u^{(s-1)}$ and $u^{(s-2)}$ terms
proportional to $m^2$ and $m$ respectively. Finding it fixes coefficients $f^{(s-2)}$ and $d^{(s-2)}$. Then
we use it to find $\J^{(s-2)}=1/m~\K^{(s-2)}+\I^{(s-2)}$, which includes only $u^{(s-2)}$ and $\chi^{(s-3)}$
terms. Going on-shell, and assuming that $\chi^{(s-3)}$ can be set to zero\footnote{For the supermultiplet
corresponding to $s=3$  ($\Ysf=7/2$) no such assumption is needed.} we can choose the $c^{(s-2)}$ coefficients
so $u^{(s-2)}=0$. This will also give $\chi^{(s-2)}=0$.

Here we repeat it for the arbitrary $q$-th level. We will show that there is a consistent choice
of parameters such that if one assumes that
all auxiliary superfields up to level $(q-1)$ vanish then $u^{(q)}$ and $\chi^{(q)}$ also vanish
\begin{equation}\label{q}
    \Bigl\{u^{(r)}=\chi^{(r)}=0~,~r=1,2,\dots,q-1\Bigr\}~\Rightarrow~u^{(q)}=0~\Rightarrow~\chi^{(q)}=0~.
\end{equation}
This result holds for all values of $q=1,2,\dots,s$ and can be used recursively. For the special case of
$q=1$, our assumption ($u^{(r)}=\chi^{(r)}=0$) is trivially satisfied because there are no such
superfields\footnote{Recall that we encapsulated this information in our convention to assign the
following values $(b_1^{(1)}=b_2^{(1)}=0)$ for an automatic drop out of these terms.}, hence the on-shell
statement $u^{(1)}=\chi^{(1)}=0$ does not rely on any hypothesis and is a pure consequence of action
(\ref{mS}) given the specific values of the parameters. Hence, the assumption for the vanishing of
$u^{(2)},~\chi^{(2)}$ is justified and so on for all higher auxiliary superfields. As a result, (\ref{q})
can be applied to prove that all auxiliary superfields indeed vanish on-shell.
Moreover, based on the $(s)$ and $(s-1)$ level results we conjecture that
\begin{equation}\label{conjecture}
    d_1^{(q)}=0~,~d_3^{(q)}=0~,~c_2^{(q)}=0~,~b_1^{(q)}=1~,~b_2^{(q)}=\frac{q}{q+1}~,~~\forall~q=1,2,\dots,s~.
\end{equation}

Our staring point is the end result of level $(q+1)$. That means that
$\J^{(q+1)}_{\a(q+1)\ad(q)}=1/m~\K^{(q+1)}_{\a(q+1)\ad(q)}+\I^{(q+1)}_{\a(q+1)\ad(q)}$
depends only on $u^{(q+1)}$ and $\chi^{(q)}$ superfields,
$\K^{(q+1)}_{\a(q+1)\ad(q)}$ has only terms proportional to $m$ and $m^2$ of the form
\begin{IEEEeqnarray*}{rl}
\K^{(q+1)}_{\a(q+1)\ad(q)}=&
m^2~\D^{\a_{q+2}}\Dd^{\ad_{q+1}}u_{\a(q+2)\ad(q+1)}
+m^2~\frac{q+2}{q+3}~\Dd^{\ad_{q+1}}\D^{\a_{q+2}}u_{\a(q+2)\ad(q+1)}\n\label{afterK(q+1)}\\[2mm]
&+m~f_1^{(q+1)}~\D^2\Dd^2u_{\a(q+1)\ad(q)}
+m~\frac{f_4^{(q+1)}}{(q+1)!}~\D_{(\a_{q+1}}\Dd^2\D^{\b}u_{\b\a(q))\ad(q)}\\[2mm]
&+m~\frac{f_5^{(q+1)}}{(q+1)!q!}
~\D_{(\a_{q+1}}\Dd_{(\ad_{q}}\D^{\b}\Dd^{\bd}u_{\b\a(q))\bd\ad(q-1))}
\end{IEEEeqnarray*}
and $\I^{(q+1)}$ is given by (\ref{Iq}) for $q\to q+1$, simplified by (\ref{conjecture}).
The coefficients $f_1^{(q+1)},~f_3^{(q+1)},~f_4^{(q+1)}$ and $d_2^{(q+1)},~d_4^{(q+1)}$ have been
determined by eliminating additional $u^{(q+2)}$ and $\chi^{(q+1)}$ terms.
Also the coefficients
$c_1^{(q+1)},~c_3^{(q+1)},~c_4^{(q+1)}$ have been fixes such that if $\chi^{(q)}=0$ on-shell then
$\J^{(q+1)}_{\a(q+1)\ad(q)}=0~\Rightarrow~u^{(q+1)}_{\a(q+1)\ad(q)}=0~\Rightarrow
~\chi^{(q+1)}_{\a(q+1)\ad(q)}=0$.

Now we use $\J^{(q+1)}$ together with $\E^{(\chi,q)}$ in order to
construct $\K^{(q)}$
\begin{IEEEeqnarray*}{l}
\K^{(q)}_{\a(q)\ad(q-1)}=
\D^{\a_{q+1}}\Dd^{\ad_{q}}\J^{(q+1)}_{\a(q+1)\ad(q)}
+\frac{q+1}{q+2}~\Dd^{\ad_{q}}\D^{\a_{q+1}}\J^{(q+1)}_{\a(q+1)\ad(q)}
+f_1^{(q)}~\D^2\Dd^2\E^{(\chi,q)}_{\a(q)\ad(q-1)}\\
\hspace{23mm}+\frac{f_4^{(q)}}{q!}~\D_{(\a_{q}}\Dd^2\D^{\b}\E^{(\chi,q)}_{\b\a(q-1))\ad(q-1)}
+\frac{f_5^{(q)}}{q!(q-1)!}~\D_{(\a_{q}}\Dd_{(\ad_{q-1}}\D^{\b}\Dd^{\bd}
\E^{(\chi,q)}_{\b\a(q-1))\bd\ad(q-2))}\\[3mm]
=m^2~\D^{\a_{q+1}}\Dd^{\ad_{q}}u_{\a(q+1)\ad(q)}
+m^2~\frac{q+1}{q+2}~\Dd^{\ad_{q}}\D^{\a_{q+1}}u_{\a(q+1)\ad(q)}
+m f_1^{(q)}\D^2\Dd^2u_{\a(q)\ad(q-1)}
\n\label{Kq}\\
+m~\frac{f_4^{(q)}}{q!}~\D_{(\a_{q}}\Dd^2\D^{\b}u_{\b\a(q-1))\ad(q-1)}
+m~\frac{f_5^{(q)}}{q!(q-1)!}
~\D_{(\a_{q}}\Dd_{(\ad_{q-1}}\D^{\b}\Dd^{\bd}u_{\b\a(q-1))\bd\ad(q-2))}\\[2mm]
+\left\{\frac{2q+1}{(q+1)^2}~d_4^{(q+1)}c_3^{(q+1)}-\frac{2q+1}{q(q+1)}f_5^{(q+1)}-f_1^{(q)}\right\}
~\D^2\Dd^2\D^{\b}\Dd^{\bd}u_{\b\a(q)\bd\ad(q-1)}\\[1mm]
+\left\{\frac{q^2}{(q+1)(q+2)}~d_4^{(q+1)}c_3^{(q+1)}-\frac{q}{q+2}f_5^{(q+1)}-\frac{q+1}{q+2}f_4^{(q)}\right\}
~\frac{1}{q!}~\D_{(\a_{q}}\Dd^2\D^{\b}\Dd^{\gd}\D^{\g}u_{\b\g\a(q-1))\gd\ad(q-1)}\\[1mm]
+\left\{-\frac{q(q-1)}{(q+1)^2}~d_4^{(q+1)}c_3^{(q+1)}+\frac{q-1}{q+1}f_5^{(q+1)}-f_5^{(q)}\right\}
~\frac{1}{q!(q-1)!}
~\D_{(\a_{q}}\Dd_{(\ad_{q-1}}\D^{\b}\Dd^{\bd}\D^{\g}\Dd^{\gd}u_{\b\g\a(q-1))\bd\gd\ad(q-2))}\\[2mm]
+\left\{2~\frac{2q+1}{q(q+1)}~d_2^{(q+1)}+2f_1^{(q)}d_2^{(q)}\right\}
~\D^2\Dd^2\D^2\chi_{\a(q)\ad(q-1)}\\[1mm]
+\left\{-\frac{2q+1}{(q+1)^2}~d_4^{(q+1)}+f_1^{(q)}d_4^{(q)}\right\}
~\frac{1}{q!}~\D^2\Dd^2\D_{(\a_{q}}\Dd^{\bd}\bar{\chi}_{\a(q-1))\bd\ad(q-1)}\\[1mm]
+\left\{-\frac{q}{q+1}~d_4^{(q+1)}+\frac{q+1}{q}f_4^{q)}d_4^{(q)}\right\}
~\frac{1}{q!}~\D_{(\a_{q}}\Dd^2\D^2\Dd^{\bd}\bar{\chi}_{\a(q-1))\bd\ad(q-1)}\\[1mm]
+\left\{-2~\frac{q-1}{q+1}~d_2^{(q+1)}+2f_5^{(q)}d_2^{(q)}\right\}
~\frac{1}{q!(q-1)!}
~\D_{(\a_{q}}\Dd_{(\ad_{q-1}}\D^{\b}\Dd^{\bd}\D^2\chi_{\b\a(q-1))\bd\ad(q-2))}\\[1mm]
+\left\{\frac{q(q-1)}{(q+1)^2}~d_4^{(q+1)}+f_5^{(q)}d_4^{(q)}\right\}
~\frac{1}{q!q!(q-1)!}
~\D_{(\a_{q}}\Dd_{(\ad_{q-1}}\D^{\b}\Dd^{\bd}\D_{(\b}\Dd^{\gd}\bar{\chi}_{\a(q-1)))\bd\gd\ad(q-2))}~.
\end{IEEEeqnarray*}
Demanding the cancellation of all terms that do not depend on mass fixes $f^{(q)}$~s as follows
\begin{IEEEeqnarray*}{l}\n\label{fq}
    f_1^{(q)}=\frac{2q+1}{(q+1)^2}~\Bigl(d_4^{(q+1)}~c_3^{(q+1)}-\frac{q+1}{q}~f_5^{(q+1)}\Bigr)~,\sn\\[2mm]
    f_4^{(q)}=\frac{q^2}{(q+1)^2}~\Bigl(d_4^{(q+1)}~c_3^{(q+1)}-\frac{q+1}{q}~f_5^{(q+1)}\Bigr)~,\sn\\[2mm]
    f_5^{(q)}=-~\frac{q(q-1)}{(q+1)^2}~\Bigl(d_4^{(q+1)}~c_3^{(q+1)}-\frac{q+1}{q}~f_5^{(q+1)}\Bigr)\sn
\end{IEEEeqnarray*}
and the $d^{(q)}$~s
\begin{equation}\label{dq}
    d_2^{(q)}=-~\frac{2q+1}{q(q+1)}~\frac{d_2^{(q+1)}}{f_1^{(q)}}~,~d_4^{(q)}=\frac{2q+1}{(q+1)^2}
    ~\frac{d_4^{(q+1)}}{f_1^{(q)}}~.
\end{equation}

The last step is to construct $\J^{(q)}=1/m~\K^{(q)}+\I^{(q)}$. The result is
\begin{IEEEeqnarray*}{l}
\J^{(q)}_{\a(q)\ad(q-1)}=m^2~u_{\a(q)\ad(q-1)}
+\left\{-4d_2^{(q)}c_1^{(q)}-d_4^{(q)}c_3^{(q)}+f_1^{(q)}\right\}~\D^2\Dd^2u_{\a(q)\ad(q-1)}
\n\label{preJq}\\[2mm]
\hspace{2.3cm}+\left\{-2d_2^{(q)}c_3^{(q)}-2d_4^{(q)}c_1^{(q)}\right\}
~\frac{1}{q!}~\D^2\Dd^{\bd}\D_{(\a_{q}}\bar{u}_{\a(q-1))\bd\ad(q-1)}\\[1mm]
\hspace{2.3cm}+\left\{-d_4^{(q)}\left(\frac{q+1}{q}~c_4^{(q)}-c_3^{(q)}\right)+f_4^{(q)}\right\}
~\frac{1}{q!}~\D_{(\a_{q}}\Dd^2\D^{\b}u_{\b\a(q-1))\ad(q-1)}\\[1mm]
\hspace{2.3cm}+\left\{-\frac{q-1}{q}~d_4^{(q)}c_3^{(q)}+f_5^{(q)}\right\}
~\frac{1}{q!(q-1)!}~\D_{(\a_{q}}\Dd_{(\ad_{q-1}}\D^{\b}\Dd^{\bd}u_{\b\a(q-1))\bd\ad(q-2))}\\[2mm]
\hspace{2.3cm}+\left\{-2d_2^{(q)}\right\}
~\frac{1}{q!(q-1)!}~\D^2\Dd_{(\ad_{q-1}}\D_{(\a_{q}}\chi_{\a(q-1))\ad(q-2))}\\[1mm]
\hspace{2.3cm}+\left\{\frac{q-1}{q}~d_4^{(q)}\right\}
~\frac{1}{q!(q-1)!}
~\D_{(\a_{q}}\Dd_{(\ad_{q-1}}\D_{\a_{q-1}}\Dd^{\gd}\bar{\chi}_{\a(q-2))\gd\ad(q-2))}~.
\end{IEEEeqnarray*}
The above dictates that if we want to make $u^{(q)}$ vanish on-shell, assuming
$u^{(r)}=\chi^{(r)}=0$\footnote{Besides dropping the last two terms of (\ref{preJq}), it also imposes
    on $u^{(q)}$ the constraint $\D^{\b}\Dd^{\bd}u_{\b\a(q-1)\bd\ad(q-2)}+
\frac{q}{q+1}\Dd^{\bd}\D^{\b}u_{\b\a(q-1)\bd\ad(q-2)}=0$.}
for $r=1,2,\dots,q-1$, we must select coefficients $c^{(q)}$~s such that:
\begin{equation}\label{cq}
    \begin{rcases*}
        4~d_2^{(q)}~c_1^{(q)}+d_4^{(q)}~c_3^{(q)}=f_1^{(q)}\\[3mm]
        d_2^{(q)}~c_3^{(q)}+d_4^{(q)}~c_1^{(q)}=0\\[3mm]
        \frac{q+1}{q}~d_4^{(q)}~c_4^{(q)}-\frac{2}{q+1}~d_4^{(q)}~c_3^{(q)}=f_4^{(q)}+\frac{q}{q+1}~f_5^{(q)}
    \end{rcases*}\Rightarrow
    \begin{cases*}
        c_1^{(q)}=-~f_1^{(q)}~\frac{d_2^{(q)}}{[d_4^{(q)}]^2-4~[d_2^{(q)}]^2}\\[3mm]
        c_3^{(q)}=f_1^{(q)}~\frac{d_4^{(q)}}{[d_4^{(q)}]^2-4~[d_2^{(q)}]^2}\\[3mm]
        c_4^{(q)}=\frac{2q}{(q+1)^2}~c_{3}^{(q)}+\frac{q}{q+1}~\frac{f_4^{(q)}}{d_4^{(q)}}
        +\frac{q^2}{(q+1)^2}~\frac{f_5^{(q)}}{d_4^{(q)}}
    \end{cases*}
\end{equation}
\section{Off-shell degrees of freedom}\label{sec:dof}
The conclusion of section (\ref{sec:Ahis}) is that the off-shell description of an arbitrary
half-integer superspin ($\Ysf=s+1/2$) supermultiplet is given by the following superspace action principle
\begin{IEEEeqnarray*}{rl}
    S_{(m,\Ysf=s+1/2)}=\hspace{-1mm}\int\hspace{-1mm} d^8z\Biggl\{\Biggr.
&~H^{\a(s)\ad(s)}\D^{\g}\Dd^2\D_{\g}H_{\a(s)\ad(s)}~+~m^2H^{\a(s)\ad(s)}H_{\a(s)\ad(s)}\n\label{fmS}\\
&-2~H^{\a(s)\ad(s)}\Dd_{\ad_s}\D^2\chi_{\a(s)\ad(s-1)}~+c.c.\\
&+\sum_{q=1}^{s}\Bigl[\Bigr.
d_2^{(q)}~\chi^{\a(q)\ad(q-1)}\D^2\chi_{\a(q)\ad(q-1)}
+c_{1}^{(q)}~u^{\a(q)\ad(q-1)}\Dd^2u_{\a(q)\ad(q-1)}~+c.c.\\
&\hspace{11mm}+d_4^{(q)}~\chi^{\a(q)\ad(q-1)}\D_{\a_q}\Dd^{\ad_q}\bar{\chi}_{\a(q-1)\ad(q)}
+c_{3}^{(q)}~u^{\a(q)\ad(q-1)}\Dd^{\ad_q}\D_{\a_q}\bar{u}_{\a(q-1)\ad(q)}\\[3mm]
&\hspace{11mm}+c_{4}^{(q)}~u^{\a(q)\ad(q-1)}\D_{\a_q}\Dd^{\ad_q}\bar{u}_{\a(q-1)\ad(q)}
+m~\left(\chi^{\a(q)\ad(q-1)}u_{\a(q)\ad(q-1)}+c.c.\right)\Bigl.\Bigr]\\
&+\sum_{q=1}^{s-1}\Bigl[\Bigr.
u^{\a(q+1)\ad(q)}\left(\Dd_{\ad_q}\D_{\a_{q+1}}\chi_{\a(q)\ad(q-1)}+
\frac{q+1}{q+2}~\D_{\a_{q+1}}\Dd_{\ad_{q}}\chi_{\a(q)\ad(q-1)}\right)\hspace{-1mm}+c.c.
\Bigl.\Bigr]
\Biggl.\Biggr\}
\end{IEEEeqnarray*}
where the coefficients $d^{(q)}$ and $c^{(q)}$ are given by the recursive relations (\ref{fq}),(\ref{dq})
and (\ref{cq}) together with the initial conditions fixed by the massless limit of the theory
\begin{equation}\label{ic}
d_2^{(s)}=-\frac{s+1}{s}~,~d_4^{(s)}=2~,~f_1^{(s)}=-1~,f_4^{(s)}=f_5^{(s)}=0~.
\end{equation}

The off-shell structure of this theory is extremely rich, since it requires the presence of two towers of
auxiliary superfields with $s$ members each of increasing rank $\{u^{(r)},~\chi^{(r)}\}~r=1,2,\dots,s$.  By
projecting the superspace action to components, we can find the component structure of the theory.  It will
include the auxiliary fields required by~\cite{Singh:1974qz,Singh:1974rc} for the off-shell description of
irreducible, arbitrary higher spins plus additional auxiliary fields required by off-shell supersymmetry.
Because all participating superfields are unconstrained  and there is no redundancy, all the components of
every superfield will participate in the off-shell component action.

It is straightforward to count the off-shell degrees of freedom of an unconstrained $(n,m)$ superfield
$\Phi_{\a(n)\ad(m)}$. The answer is $16(n+1)(m+1)$
bosons\footnote{For details see~\cite{Gates:2017hmb}.} and equal number of fermions. An  exception
is the case were we can impose a reality condition
($\Phi=\bar{\Phi}~\Rightarrow~n=m$).
In that case, the real superfield $\Phi_{\a(n)\ad(n)}$ carries $8(n+1)^2$ bosons and equally many fermions.
Therefore the total number of off-shell degrees of freedom of this theory is
\begin{equation}
    8(s+1)^2+2\times\sum_{n=1}^{s}16(n+1)n=\frac{8}{3}~(s+1)(4s^2+11s+3)~.
\end{equation}
\section{Summary}
The supersymmetric Fierz-Pauli program of constructing superspace Lagrangians for higher spin supermultiplets
has been a long standing question since the birth of supersymmetry. In this paper we answer this question
for arbitrary half-integer ($\Ysf=s+1/2$) supermultiplets that on-shell describe the propagation of free
massive spins $j=s+1,~j=s+1/2,~j=s+1/2,~j=s$. We find that the off-shell superspace action description of this
supermultiplet requires a tower of pairs of auxiliary superfields $u_{\a(q)\ad(q-1)},~\chi_{\a(q)\ad(q-1)}$
with $q=1,2,\dots,s$ and it has the form \eqref{fmS}.

The coupling constants $d_2^{(q)},~d_4^{(q)}$ and $c_1^{(q)}, c_3^{(q)}, c_4^{(q)}$ of \eqref{fmS}
are given by the recursive relations \eqref{fq}, \eqref{dq} and \eqref{cq} with initial conditions \eqref{ic}.
They have been determined such that on-shell (\emph{i}) all auxiliary superfields vanish
($u_{\a(q)\ad(q-1)}=\chi_{\a(q)\ad(q-1)}=0$) (\emph{ii}) we
generate the appropriate constraints for superfield $H_{\a(s)\ad(s)}$ ($\D^{\a_s}H_{\a(s)\ad(s)}=0,~\Box
H_{\a(s)\ad(s)}=m^2H_{\a(s)\ad(s)}$) which allow only the above mentioned spin degrees of
freedom to propagate and (\emph{iii}) the massless limit of the action gives a smooth transition to the correct description
of the arbitrary massless half-integer superspin supermultiplet by decoupling all auxiliary
superfields with the exception of $\chi_{\a(s)\ad(s-1)}$ which becomes the compensator of $H_{\a(s)\ad(s)}$,
as required by the gauge redundancy of the massless theory.

These recursion relations can be iterated
systematically to extract the numerical value of all coefficients. For example, as demonstrated by
\eqref{c(s)}, \eqref{c(s-1)} and \eqref{ic},~\eqref{d(s-1)},~\eqref{conjecture} the first two levels of
coefficients, for arbitrary values of the parameter $s$, are:
\begin{equation}\label{rc1}
    c_1^{(s)}=\frac{1}{4}~\frac{s(s+1)}{2s+1},~
    c_3^{(s)}=\frac{1}{2}~\frac{s^2}{2s+1},~
    c_4^{(s)}=\frac{s^3}{(s+1)^2(2s+1)},~
    d_2^{(s)}=-\frac{s+1}{s},~d_4^{(s)}=2
\end{equation}
\begin{IEEEeqnarray*}{l}\n\label{rc2}
c_1^{(s-1)}=\frac{1}{16}~\frac{(s+1)s(s-1)(2s-1)}{(2s+1)^2},~
c_3^{(s-1)}=-~\frac{1}{8}~\frac{s(s-1)^2(2s-1)}{(2s+1)^2},~\\
c_4^{(s-1)}=\frac{1}{4}~\frac{(s-1)^3}{s(2s+1)},~
d_2^{(s-1)}=\frac{(s+1)(2s+1)}{s^2(2s+1)},~
d_4^{(s-1)}=2~\frac{2s+1}{s^2}
\end{IEEEeqnarray*}
For the special case of $s=1$, \eqref{rc1} give the numerical value of the coefficients required for the
description of massive $\Ysf=3/2$ supermultiplet and they are in agreement with the findings of
\cite{Gates:2013tka}. For $s=2$
\eqref{rc1}, \eqref{rc2} give the numerical value of coefficients required for the description of massive
$\Ysf=5/2$ supermultiplet.

Moreover, the coefficients $d_1^{(q)}$ and $ d_3^{(q)}$ vanish for all values of $q$.
For $q=s$ this is a consequence of the gauge invariance of the massless action and it's effect trickles
down to all other levels. As a result $c_2^{(q)}$ drops out of all equations
(for example \eqref{c2(s-1)}), remains undetermined
and not relevant for the on-shell spectrum of the theory. For simplicity we set it to zero.

A characteristic feature of the theory is that it requires the presence of the `bare' superfield
$\chi_{\a(s)\ad(s-1)}$. This means that the constrained superfield
($\Gamma_{\a(s-1)\ad(s-1)}\propto\Dd^{\ad_s}\bar{\chi}_{\a(s-1)\ad(s)}$) approach of
\cite{Kuzenko:1993jp,Kuzenko:1993jq}
can not be used to generate this result. From the view point of
the massive theory the unconstrained superfield approach of \cite{Gates:2013rka,Gates:2013ska} seem to be the
more appropriate variables that one should consider.

Due to the plethora of the auxiliary superfields, the off-shell structure of the theory is extremely rich and
the number of off-shell degrees of freedom scale as $\sim s^3$. Specifically, the theory carries
$\frac{8}{3}(s+1)(4s^2+11s+3)$ bosons and equal number of fermions. In contrast, the on-shell degrees of
freedom are just the $4(s+1)$ polarizations of bosonic spins $j=s+1,~j=s$ and an equal number of states
coming from the two $j=s+1/2$ fermionic spins.
\section*{Acknowledgments} This work is supported in part by S.~James Gates Jr.'s
endowment of the Ford Foundation
Professorship of Physics at Brown University and the author gratefully acknowledge the support of the Brown
Theoretical Physics Center.
\begin{multicols}{2}
    {\small \bibliographystyle{hephys} \bibliography{references}}
\end{multicols}
\end{document}